\documentclass[]{aiaa-tc}
\usepackage{hyperref}
\hypersetup{
	colorlinks=true,
	linktoc=all,
	linkcolor=red,
	citecolor=blue,
}
\usepackage{amsmath,amssymb,stmaryrd,array} 
\usepackage{epsfig,epstopdf,color,soul,rotating,overpic,psboxit}
\usepackage{subfig} 
\usepackage{array}

\newcommand{\fig}{Fig.~}
\newcommand{\figs}{Figs.~}
\newcommand{\eq}{Eq.~}
\newcommand{\eqs}{Eqs.~}

\def\drawline#1#2{\raise 2.5pt\vbox{\hrule width #1pt height #2pt}}
\def\spacce#1{\hskip #1pt}

\def\bdash{\hbox{\drawline{4}{.5}\spacce{2}}}
\def\dashed{\bdash\bdash\bdash\hbox{\drawline{4}{.5}}\nobreak\ }

\definecolor{red}{rgb}{0.8500, 0.1250, 0.0480} 
\definecolor{blue}{rgb}{0, 0.4470, 0.7410}
\definecolor{green}{rgb}{0.4660, 0.6740, 0.1880}
\definecolor{gray}{rgb}{0.7, 0.7, 0.7}

\title{Effects of wall-normal and angular momentum injections in airfoil separation control}

 \author{
  Phillip M. Munday\thanks{Graduate Research Assistant, Department of Mechanical Engineering, pmm06d@my.fsu.edu, Student Member AIAA; Current Affiliation: Odyssey Systems Consulting.}
  \ and Kunihiko Taira\thanks{Associate Professor, Department of Mechanical Engineering, ktaira@fsu.edu, Associate Fellow AIAA.}\\
  {\normalsize\itshape
   Department of Mechanical Engineering, Florida State University, Tallahassee, FL 32310, USA}
 }

\AIAApapernumber{YEAR-NUMBER}
\AIAAconference{Conference Name, Date, and Location}
\AIAAcopyright{\AIAAcopyrightD{YEAR}}

\begin{document}

\maketitle 


\begin{abstract}

The objective of this computational study is to quantify the influence of wall-normal and angular momentum injections in suppressing laminar flow separation over a canonical airfoil.  Open-loop control of fully separated, incompressible flow over a NACA 0012 airfoil at $\alpha = 9^\circ$ and $Re = 23,000$ is examined with large-eddy simulations. This study independently introduces wall-normal momentum and angular momentum into the separated flow using swirling jets through model boundary conditions.  The response of the flow field and the surface vorticity fluxes to various combinations of actuation inputs are examined in detail.  It is observed that the addition of angular momentum input to wall-normal momentum injection enhances the suppression of flow separation.  Lift enhancement and suppression of separation with the wall-normal and angular momentum inputs are characterized by modifying the standard definition of the coefficient of momentum.  The effect of angular momentum is incorporated into the modified coefficient of momentum by introducing a characteristic swirling jet velocity based on the non-dimensional swirl number.  With this single modified coefficient of momentum, we are able to categorize each controlled flow into separated, transitional, and attached flows.  
\end{abstract}

\section*{Nomenclature}

\vspace{-0.1in}
{\small 
\noindent\begin{tabular}{@{}lcl@{}}
$A$             &=& Planform area\\
$A_j$           &=& Actuator jet area\\
$c$             &=& Chord length \\
$C_D$           &=& Coefficient of drag \\
$C_L$           &=& Coefficient of lift \\
$C_\mu$         &=& Coefficient of momentum\\
$C_\mu^*$       &=& Modified coefficient of momentum\\
$F_x$           &=& Drag force \\
$F_y$           &=& Lift force \\
$G_n$           &=& Wall-normal momentum flux\\
$G_\theta$      &=& Tangential momentum flux\\
$l_z$           &=& Spanwise extent\\
$M_\infty$      &=& Freestream Mach number\\
$\hat{\boldsymbol{n}}$ &=& Wall-normal unit vector\\
$N_\text{act}$  &=& Number of actuators\\
$p$             &=& Pressure\\
$\overline{p}$  &=& Time-average pressure\\
$p_\infty$      &=& Freestream pressure\\
$Q$             &=& Q criteria\\
$r$             &=& Radial direction\\
$r_a$           &=& Radius of actuator\\
\end{tabular} 
}

{\small 
\noindent\begin{tabular}{@{}lcl@{}}
$Re$            &=& Reynolds number\\
RMS             &=& Root mean square\\
$S$             &=& Swirl number\\
$t$             &=& Time\\
$\boldsymbol{u}=(u_x,u_y,u_z)$  &=& Velocity (streamwise, vertical, spanwise)\\
$\boldsymbol{u}'=(u'_x,u'_y,u'_z)$  &=& Velocity fluctuation (streamwise, vertical, spanwise)\\
$\overline{\boldsymbol{u}}=(\overline{u}_x,\overline{u}_y,\overline{u}_z)$  &=& Time-average velocity (streamwise, vertical, spanwise)\\
$U_c$           &=& Convective velocity\\
$u_j$           &=& Characteristic jet velocity\\
$u_j^*$         &=& Modified characteristic jet velocity\\
$u_{n}$         &=& Wall-normal actuator velocity\\
$u_{\theta}$    &=& Azimuthal/rotational actuator velocity\\
$u_\tau$        &=& Friction velocity\\
$U_\infty$      &=& Freestream velocity\\
$\boldsymbol{x} = (x,y,z)$ &=& Spatial coordinates (streamwise, vertical, spanwise)\\
$\boldsymbol{x}^+= (x^+,y^+,z^+)$ &=& Spatial coordinates in wall units\\
\end{tabular} 
}
~\\

\noindent Greek symbols\\
{\small
\noindent\begin{tabular}{@{}lcl@{}}
$\alpha$        &=& Angle of attack, deg \\
$\Gamma_n$      &=& Wall-normal circulation of actuator jet\\
$\nu$           &=& Kinematic viscosity\\
$\rho_\infty$   &=& Freestream density\\
$\boldsymbol{\Sigma}$   &=& Diffusive vorticity flux\\
$(\sigma_x,\sigma_y,\sigma_z)$ &=& Time-average wall-normal diffusive vorticity flux (streamwise, vertical, spanwise)\\
$\tau_{xy}$     &=& Spanwise Reynolds stress\\
$\boldsymbol{\omega} = (\omega_x,\omega_y,\omega_z)$   &=& Vorticity (streamwise, vertical, spanwise)\\
\end{tabular} 
}

\section{Introduction}

Separated flow is a major cause for degradation of airfoil performance at high angles of attack. Oncoming flow encountering such an airfoil must maneuver around the leading edge, resulting in an adverse pressure gradient along the suction surface.  Beyond a certain angle of attack, the adverse pressure gradient causes the boundary layer to separate from the wing surface. Such behavior of the boundary layer can be avoided by increasing momentum in the streamwise direction of the flow in order to oppose the adverse pressure gradient.  Flow control is a technique that has been employed to suppress separated flow and mitigate accompanying detrimental effects [\citen{Lachmann_61, Chang76}].  The additional streamwise momentum can be introduced to the boundary layer either directly or by utilizing freestream momentum to energize the boundary layer to avoid separation [\citen{Gad-el-Hak00}]. Direct addition of momentum to the boundary layer energizes the flow for reattachment.  Alternatively, one can consider pulling the freestream momentum towards the suction surface by inducing or enhancing mixing between the freestream and the boundary layer. 

As a means to prevent or delay separation, active and passive flow control actuators can be utilized to introduce perturbations to the flow field [\citen{Gad-el-Hak00,Cattafesta:PAS08, Joslin09, Cattafesta:ARFM11}]. Active flow control is defined by the addition of external energy to the flow, and can be performed with an assortment of flow control actuators [\citen{Cattafesta:ARFM11}], including steady and unsteady blowing/suction [\citen{ChenAIAA13, Lachmann_61, Wu:JFM98}], pulsed jets [\citen{Hipp:AIAAJ17}], synthetic jets [\citen{Glezer:ARFM02}], sweeping jets [\citen{VastaAIAA12, SeeleAIAA12}], plasma actuators [\citen{GreenblattAIAA08, Corke_ARFM2010}], thermal actuators [\citen{Yeh:AIAA17, Yeh:TFEC17}], and vortex generator jets [\citen{Compton_AIAA92, Selby_EF92, Zhang:PF03, GrossAIAA10}].  Passive flow control devices modify the flow without external energy input and include, leading edge modification [\citen{Pedro_AIAA2008, Skillen:AIAAJ15}], vortex generators [\citen{Kerho:JA93, Lin:PA02}], and riblets [\citen{BechertAIAA97,  Choi:JFM93}]. The flow control actuators listed above do not encompass all devices that have been developed, but provides an idea of the extent of the variety of actuators considered in previous flow control studies.  

The actuators mentioned above have been implemented experimentally and numerically in different scenarios to modify separated flows around canonical airfoils.  Such efforts have relied mostly on experiments as the studies have often been closely related to actuator developments.  Let us consider the effectiveness of flow control for separated flows over canonical airfoils at moderate Reynolds numbers.  Seifert and Pack [\citen{Seifert:AIAA99}] performed experiments with synthetic jets for flow over NACA 0012 and 0015 airfoils at Reynolds numbers $Re$ of $1.5\times 10^6$ to $23.5\times10^6$ and Mach numbers $M_\infty$ of $0.2$ to $0.65$.  At post-stall angles of attack ($\alpha \gtrsim 8^\circ$), applying control can reattach the flow with an approximate $50\%$ increase in lift and $50\%$ decrease in drag. Pulsed jets have also been utilized recently by Hipp et al.~[\citen{Hipp:AIAAJ17}] to reattach flow separated from the leading edge of an NACA $64_3$-618 airfoil at $Re = 64,000$.

Similar efforts have also been undertaken by plasma actuators.  For example, Greenblatt et al.~[\citen{GreenblattAIAA08}] examined the use of dielectric barriar discharge (DBD) plasma actuators to control separated flow over flat-plate and Eppler E338 airfoils at $3000 < Re < 20,000$ with micro air vehicle application in mind.  In another study, Little et al.~[\citen{LittleEF10}] used a dielectric barrier discharge plasma actuator to modify the separation due to the deflection flap of a high-lift airfoil at $Re$ of $\mathcal{O}( 10^5$). 
Effects of control, which include delaying separation and reducing the size of the separated region, are dependent on the spatial and temporal waveforms of the control input. 

In the same spirit, using micro-vortex generators, Lin et al.~[\citen{LinJA94}] reduced flap separation over a three-element airfoil in high-lift configuration. By mitigating separation, the lift-to-drag ratio was doubled in their study. More recently, Rathay et al.~[\citen{RathayAIAAJ14b, RathayAIAAJ14a}] and Seele et al.~[\citen{SeeleAIAA12, SeeleAIAA13}] increased side force on a scaled vertical tail stabilizer using synthetic jet actuators and sweeping jet actuators. The former studies laid the foundation for a successful full-scale flow control on a vertical tail using sweeping jet actuators [\citen{WhalenAIAA15}] and their implementation on the Boeing 757 ecoDemonstrator. The effectiveness of the sub-boundary layer vortex generators is highly dependent on the device geometry, spacing, height, and angle.  With the aforementioned control approaches, flow separation is mitigated with different actuators but often leveraging similar control mechanisms. 

While challenging and computationally expensive, numerical simulations have also been performed to study flow control on NACA airfoils. The complex three-dimensional nature of separated flow under the influence of control leads to the emergence of a wide range of spatial scales in the flow field. Hence, computational analysis of flow control at moderate Reynolds number requires substantial resources to perform a sizable number of parameter study with high fidelity.  In addition to resolving the baseline conditions at Reynolds numbers similar to experiments, replicating an actuator introduces added complexity [\citen{RajuIJFC09}]. Earlier numerical studies have examined the effectiveness of blowing/suction [\citen{Wu:JFM98, HuangJA04}] and vortex generators [\citen{Shan:CF08}].  More recently, high-fidelity, three-dimensional large-eddy simulations (LES) of complex interactions between the flow over an airfoil and plasma actuator based forcing (nsDBD) as well as synthetic jet actuation have been performed by Mohan and Gaitonde [\citen{MohanJA17}] and You et al. [\citen{You:PF08}], respectively, at $Re = 10^5$ and $8.96\times 10^5$. These types of complex interactions were further investigated by Abe et al.~[\citen{AbeAIAA13}] at $Re = 6.3\times 10^4$, in which the perturbations with different spanwise wavelengths were considered to model internal effects of the synthetic jet.  Based on this study, Sato et al.~[\citen{SatoAIAA15}] modeled the effects of plasma actuators and performed a parametric sweep to determine the optimal control settings for separation mitigation. These studies highlight the challenge in replicating and modeling the influence of specific actuator inputs.

Regardless of the selected type of actuator, we can view the flow to be affected by means of mass, momentum, and energy injections.  Active flow control actuators can in an obvious manner add the aforementioned forcing inputs. Passive flow control actuators also introduce those perturbations in response to the flow negotiating the existence of the actuators.  Instead of replicating flow control inputs for each and every type of actuator, we aim to understand the response of the separated flow to fundamental control inputs that these actuators can introduce.  Herein, we particularly focus on controlled cases in which laminar separation from the leading edge can be modified by altering the transition dynamics to enhance mixing downstream and reattach the flow.  The present paper in particular investigates the influence of steady wall-normal momentum and angular momentum injections for separated flow over a NACA 0012 airfoil at $\alpha = 9^\circ$ for  $Re = 23,000$.  This particular flow condition for the present parametric flow control study is chosen due to the availability of computational resources and the benchmark data [\citen{Kojima:JA2013}].  The three-dimensional LES herein sheds light on the separated flow physics and identifies the key actuation mechanism for reattaching the flow at a high angle of attack. 

The remainder of the paper is organized as follows. In section \ref{sec:SimMeth}, we present the computational approach and the flow control setup. Section \ref{sec:FlowPhys} details the influence of control on the separated flow, with focus on vortex dynamics. With the understanding of the influence of control perturbations on the flow, we quantify the effects of control inputs in section \ref{sec:Quantification} and examine the resulting change in aerodynamic forces. We in particular discuss the quantification of the lift change due to the actuation by modifying the coefficient of momentum to incorporate the effects of swirl injection. We end the paper by offering some concluding remarks.


\section{Approach}
\label{sec:SimMeth}

\subsection{Simulation setup}

We perform LES of flow over a NACA 0012 airfoil using an incompressible flow solver, {\tt Cliff} ({\tt CharLES} software package, Cascade Technologies [\citen{Ham:2004CTR, HAM:CTR06}]). The incompressible Navier--Stokes equations are discretized using second-order finite-volume and time-integration schemes. Incorporating energy conservation properties [\citen{Morinishi_JCP1998}], the solver is capable of handling structured and unstructured grids.  In the present study, we define the Reynolds number as $Re = U_\infty c/\nu = 23,000$, which is based on the freestream velocity, $U_\infty$, the chord length of the airfoil, $c$, and the kinematic viscosity, $\nu$. To predict the flow field at this Reynolds number, eddy viscosity is introduced with the Vreman subgrid-scale model [\citen{Vreman:2004PF}]. The simulations consider fully separated flow at $\alpha = 9^\circ$ and $Re = 23,000$ for the purpose of mitigating separation and achieving both lift increase and drag reduction. 

\begin{figure}
\begin{center}  
\includegraphics[width=0.93\textwidth]{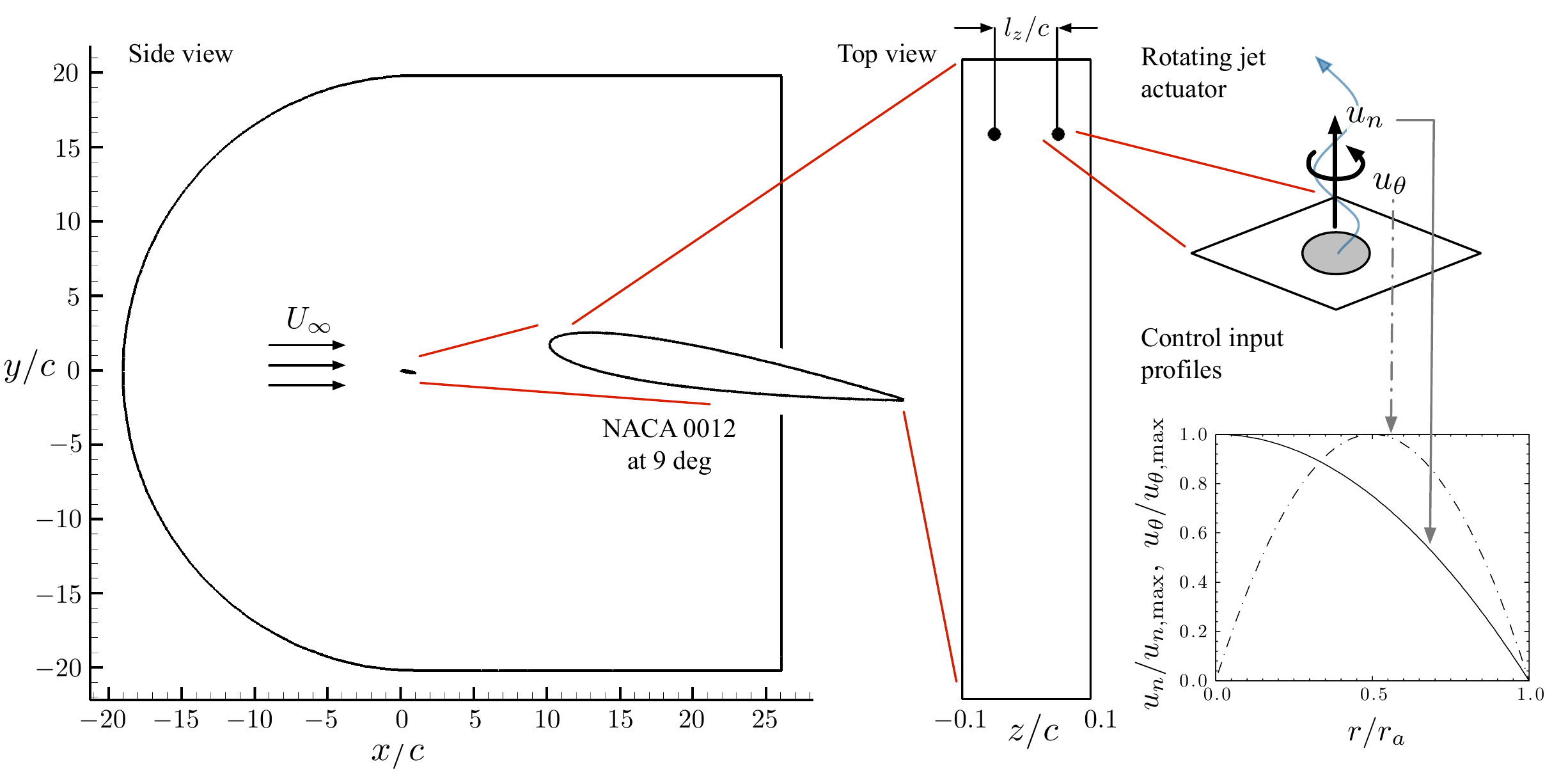}
\caption{Computational domain (left) and the actuator setup (middle and right) considered in the present LES. The insert in the side view diagram depicts the NACA 0012 airfoil at $\alpha = 9^\circ$.  A top down view in the middle shows the airfoil with two actuators over the spanwise extent (right).}
\label{fig:domain}
\end{center}  
\end{figure}

The overall size of the computational domain is $(x/c,y/c,z/c) \in [-19,26]\times [-20,20]\times [-0.1,0.1]$, as illustrated in \fig \ref{fig:domain}.  The spatial directions $x$, $y$, and $z$ refer to the streamwise, vertical, and spanwise directions, respectively. At the inlet, uniform flow of $\boldsymbol{u}/U_\infty = (1,0,0)$ is prescribed. Stress-free boundary conditions are applied at the far-field (top and bottom) boundaries.  A convective outflow condition, $\frac{\partial \boldsymbol{u}}{\partial t} + U_c \frac{\partial\boldsymbol{u}}{\partial x} = 0$, is prescribed at the outlet.  Here, the convective velocity ($U_c$) is the time-average, boundary-normal velocity component at the outlet. This boundary condition allows wake structures to leave the domain without disturbing the near-field solution. In the spanwise direction, the flow is taken to be periodic.  The spanwise extent of the present model problem is set to $0.2c$, which should capture the dominant flow physics.  However, instabilities and spanwise variations with spanwise wavelengths greater than $0.2c$ cannot be analyzed in the current study.  Such investigation may be possible with emerging approaches [\citen{Schmid:PRF17}] that can examine the influence of spanwise effects without significant computational burden.  Future efforts will examine the influence of large spanwise extents, but are beyond the coverage of the present investigation.

The present study utilizes a hybrid structured/unstructured spatial discretization. A structured grid is used to achieve adequate resolution in the near field of the airfoil and an unstructured far-field grid is utilized to reduce the total number of points throughout the whole computational domain. The baseline mesh is planar (two-dimensional) and is extruded in the spanwise direction. In the region with an unstructured grid, the triangular cells discretize the $x$-$y$ plane and form triangular prisms through spanwise extrusion.  Additional refinement is introduced in the vicinity of the actuators in order to resolve the fine-scale flow structures produced by the interaction of the incoming flow and the actuator inputs. Based on the domain size and spatial resolution, the resulting computational domain is comprised of approximately $40\times10^6$ grid points.

To perform wall-resolved LES in the present investigation, sufficient resolution is necessary near the airfoil surface. The largest non-dimensional wall spacing along the suction surface of the airfoil is $\Delta x^+ \approx 6$, $\Delta y^+ \approx 0.5$, and $\Delta z^+ \approx 12$ ($\boldsymbol{x}^+ \equiv u_\tau \boldsymbol{x}/\nu$, where $u_\tau$ is the friction velocity) [\citen{Choi:PF12}]. No-slip, no-penetration boundary conditions are applied on the airfoil surface. The surface mesh of the airfoil is refined such that 30 points span across each actuator. For the controlled cases, an actuator velocity profile, described in detail in Section \ref{sec:control}, is specified at the actuator locations on the surface of the airfoil. As shown in \fig \ref{fig:domain}, the spanwise extent of the computational domain is $l_z/c = 0.2$ with two equally spaced actuators on the surface. This setup yields an actuator spacing of $(l_z/c)/N_\text{act} = 0.1$, which is used for this study. 

Throughout this paper, the time-average lift, drag, and pressure coefficients are defined as 
$C_L \equiv {F_L}/({\frac{1}{2} \rho_\infty U_{\infty}^2A})$, 
$C_D \equiv {F_D}/({\frac{1}{2} \rho_\infty U_{\infty}^2A})$, and 
$C_P \equiv {\overline{p} - p_{\infty}}/({\frac{1}{2} \rho_\infty U_{\infty}^2})$,
where $A = l_z c = 0.2c^2$ is the planform area of the airfoil and $\rho_\infty$ is the freestream density. The time-average forces, $F_L$ and $F_D$, represent lift and drag, respectively. The coefficient of pressure is offset by the free-stream pressure value $p_\infty$ and is normalized by the dynamic pressure.  To produce accurate time-averaged flow data, all of the cases considered a time greater than 40 is sufficient for the transients to subside.  To ensure initial transients are flushed out of the computational domain, averaging is performed from $t=45$ and is averaged until a time of at least $t=150$.

The computational requirements for the computational domain size, boundary condition treatment, and resolution are examined with care. To ensure the validity of our computational approach, we compare our computational results for the baseline flow with those reported in Kojima et al.~[\citen{Kojima:JA2013}] and Yeh et al.~[\citen{Yeh:AIAA17}].  Upon examining the flow field, airfoil pressure distribution, and forces, we observe good agreement as they have been reported in detail in our prior publication [\citen{MundayAIAA14}].  Here, we show the comparison of lift and drag forces and the pressure distribution from the present study and those from Kojima et al.~[\citen{Kojima:JA2013}] as well as Yeh et al.~[\citen{Yeh:AIAA17}] as part of the validation. We note in passing that Yeh et al.'s results are from a compressible flow at the same Reynolds number but with $M_\infty = 0.3$.  With the computational approach being validated, we consider the applications of flow control to mitigate separation in what follows. 

\begin{figure}
\begin{center}
\vspace{4mm}
\includegraphics[width=0.4\textwidth]{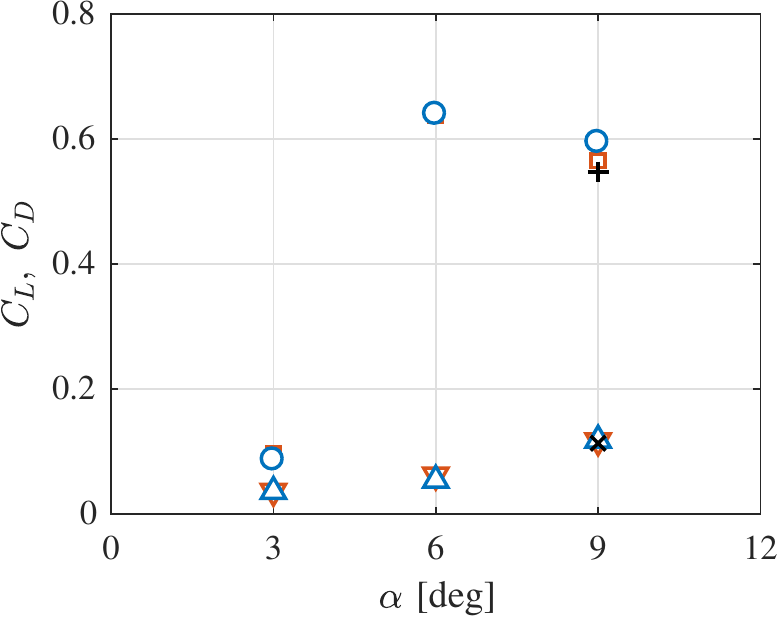}
\includegraphics[width=0.4\textwidth]{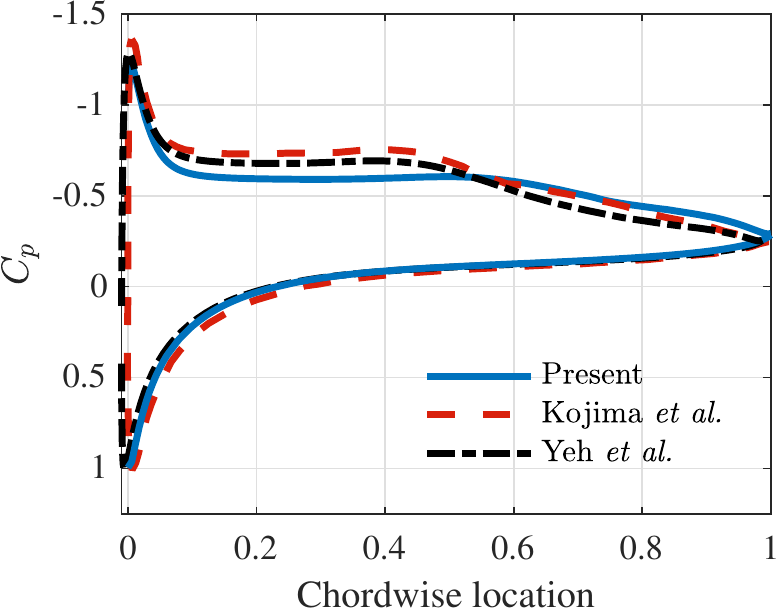}
\caption{(Left) Comparison of lift ({\Large \color{blue}$\circ$}, {\small \color{red}$\square$}, {\small \color{black}$+$}) and drag coefficients ({\small \color{blue}$\triangle$}, {\small \color{red}$\nabla$}, {\small \color{black}$\times$}). (Right) Comparison of pressure distribution over the airfoil.  For both plots, data are shown from present study (blue), Kojima et al. [\citen{Kojima:JA2013}] (red), and Yeh et al.~[\citen{Yeh:AIAA17}] (black). } 
\label{fig:validation}
\end{center}  
\end{figure}


\subsection{Actuation setup}
\label{sec:control}

We introduce steady circular jets with swirl on the top (suction-side) surface of the airfoil specified by velocity boundary conditions.  A canonical setup for the actuators is shown in \fig \ref{fig:domain}.  The actuator jets are specified over the circular regions with a radius of $r_a/c = 0.01$ placed at $10 \%$ chord location, shown by the black circles in \fig \ref{fig:domain} (middle).  This location, which is near the natural separation point.

The present problem setup is chosen to assess the influence of wall-normal and angular momentum injections independently by prescribing the wall-normal ($u_n$) and azimuthal ($u_\theta$) velocity profiles as
\begin{equation}
   \frac{u_n}{u_{n,\max}} = 1 - \left(\frac{r}{r_a}\right)^2 , 
   \quad
   \frac{u_\theta}{u_{\theta,\max}} = 4\left(1 -\frac{r}{r_a} \right)\frac{r}{r_a},
   \label{eq:vel_profiles}
\end{equation}  
which are shown in \fig \ref{fig:domain} (right). We consider arrangements where swirl is introduced in a co-rotating (COR) or counter-rotating (CTR) manner.  The magnitude of the wall-normal velocity is chosen such that the coefficient of momentum is on the same order of magnitude as in previous successful separation control studies [\citen{GreenblattEAE15}]. Moreover, the magnitude of azimuthal velocity is selected such that the maximum velocity is on the same order of magnitude as the freestream, which can be regarded as the approximate upper bound for rotational motion induced by vortex generators. 

\begin{figure}
\begin{center}
\includegraphics[width=0.4\textwidth]{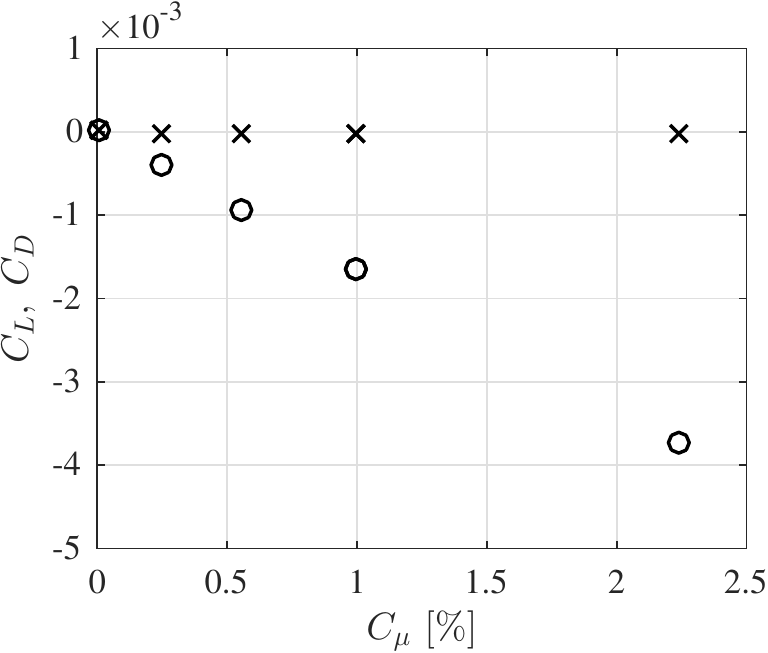}
\caption{Lift ($\circ$) and drag ($\times$) forces exerted on the airfoil by injecting wall-normal momentum in otherwise quiescent flow ($U_\infty = 0$).  Values are reported per actuator.}
\label{fig:Force_Act}
\end{center}
\end{figure}

While the above azimuthal velocity input $u_\theta$ in \eq (\ref{eq:vel_profiles}) adds rotational momentum from the actuator location, it should be noted that the profile does not generate wall-normal circulation 
$\Gamma_n$.  That is
\begin{equation}
   \Gamma_n = \int_{0}^{r_a} \omega_n (2\pi r) {\rm d}r 
   = \int_{0}^{r_a} \frac{1}{r}\frac{\partial}{\partial r} (r u_\theta) (2\pi r) {\rm d}r
 	\equiv 0,
\end{equation}
which is in general true for any velocity profiles of $u_\theta \propto (r/r_a)^\beta (1-r/r_a)^\gamma$, given $\beta$ and $\gamma >0$.

To examine the fundamental influence of the above actuation inputs, \eq (\ref{eq:vel_profiles}), on separated flow over the airfoil, we consider a large number of flow control cases.  A summary of all the controlled cases are tabulated in Appendix \ref{app:A9}.  As we shall see in later discussions, we initially report the control effort in terms of the non-dimensional coefficient of momentum [\citen{PoissonQuinton48, PoissonQuinton61}] 
\begin{equation}
 C_\mu = \frac{\rho_\infty u_{n,\max}^2 A_j N_\text{act}}{\frac{1}{2}\rho_\infty U^2_{\infty} A}, 
 \label{eq:Cmu}
\end{equation} 
in which the variables $A_j = \pi r_a^2$ and $N_\text{act}$ are the jet area and number of actuators, respectively. In the above definition, the maximum actuation velocity has been adopted.  One can also use the spatially averaged velocity as the characteristic jet velocity instead.  We later discuss how the rotational input can be incorporated into the traditional coefficient of momentum by modifying the definition to better quantify the influence of two independent control inputs through a single non-dimensional parameter. Throughout this study, we keep our actuation inputs to be steady to limit the numbers of parameters to consider.  In the Appendix, all of the control input values are listed along with the resulting force coefficients.

In the present study, the forces introduced by the actuators are included in the reported lift and drag on the airfoil. To quantify the amount of force exerted on the airfoil by an actuator with wall-normal injection, we consider evaluating the lift and drag induced by the actuators in otherwise quiescent condition ($U_\infty = 0$). These forces are found to be relatively small as shown in \fig \ref{fig:Force_Act}.  The largest $C_\mu$ of $2.2\%$ seen in the figure introduces a negative lift that is less than $1.5\%$ in magnitude of the baseline lift. Moreover, we see essentially no drag being induced solely by the actuation.  The changes in the lift and drag achieved with flow control as we report later are significantly larger in magnitude compared to the values seen in \fig \ref{fig:Force_Act}.


\section{Flow physics}
\label{sec:FlowPhys}

\subsection{Baseline flow}

\begin{figure}
\begin{center}
{\small
\begin{tabular}{m{0.125\textwidth}m{0.35\textwidth}m{0.35\textwidth}m{0.02\textwidth}} \hline 
\center{ \parbox{0.125\textwidth}{Case \\ $C_\mu$ \\ $u_{\theta,\max}/U_\infty$}} &
\center{ \parbox{0.35\textwidth}{ {Instantaneous ($Q$, $C_p$)} \\ {Time-averaged ($\overline{u}_x=0$, $\tau_{xy}$)}}} & 
\center{ Surface vorticity flux }  \parbox{0.01\textwidth}{ {$\sigma_{x}$}  \\ $\sigma_{y}$ \\ {$\sigma_{z}$}} & \\  \hline
 \parbox{0.08\textwidth}{Baseline\\ $0$ \\ $0$}& \parbox{0.325\textwidth}{\includegraphics[width=0.325\textwidth]{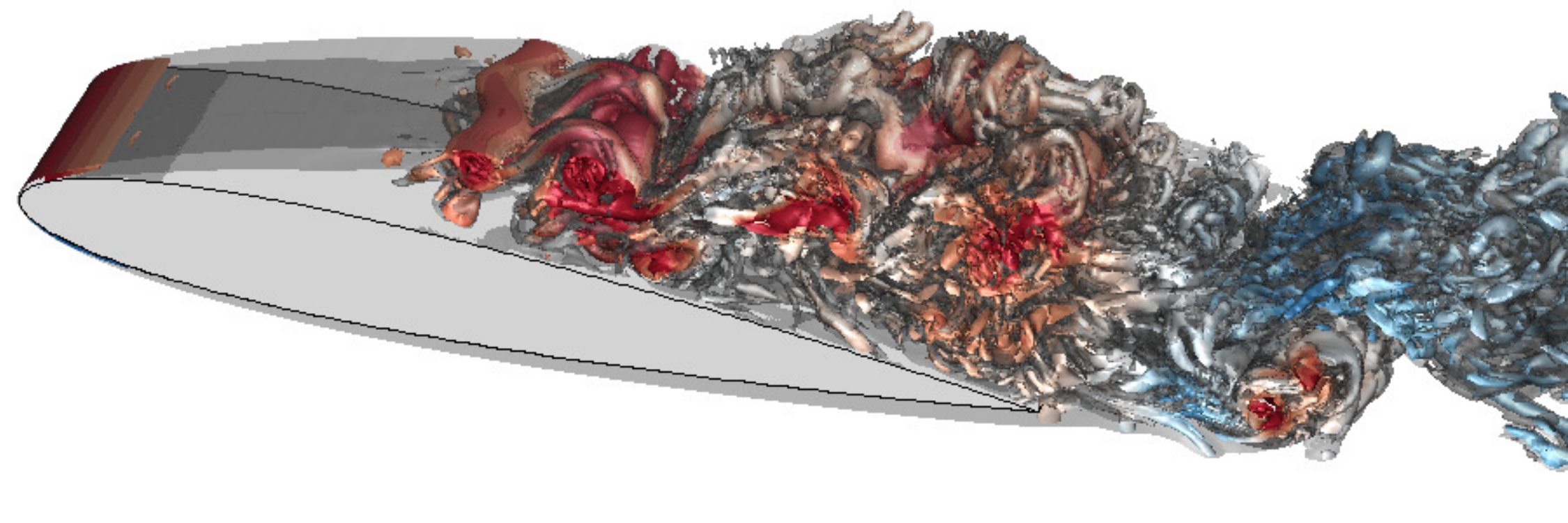}\\   
 \includegraphics[width=0.325\textwidth]{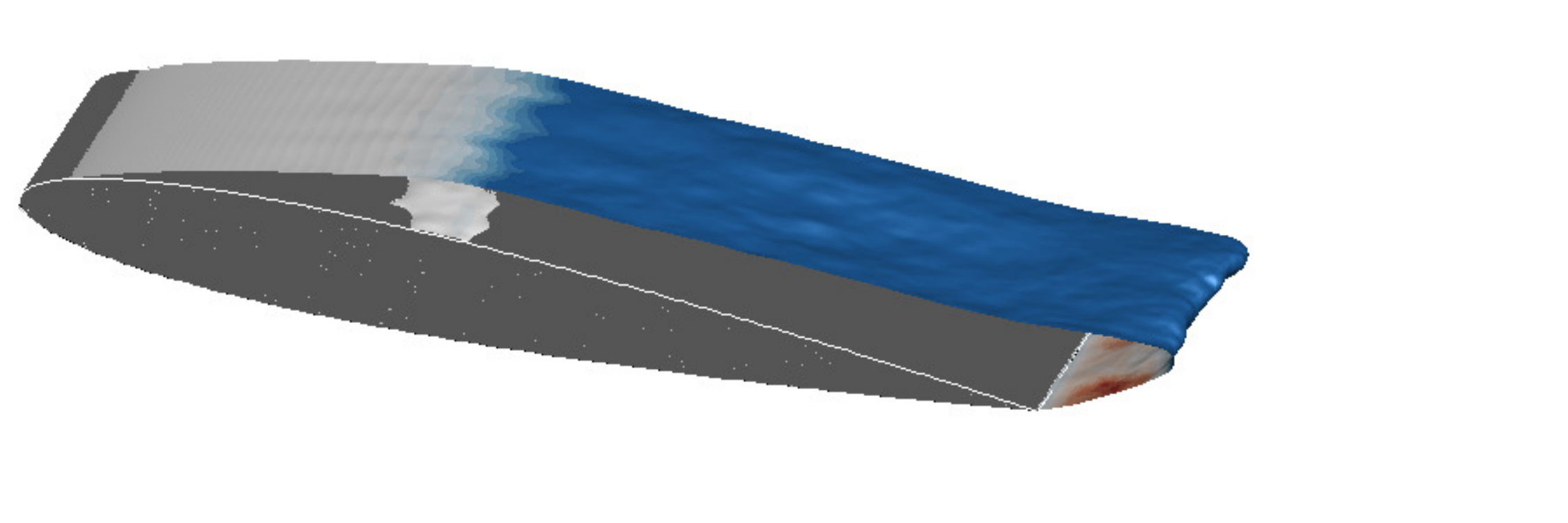}} & 
 \parbox{0.08\textwidth}{\frame{\includegraphics[width=0.35\textwidth]{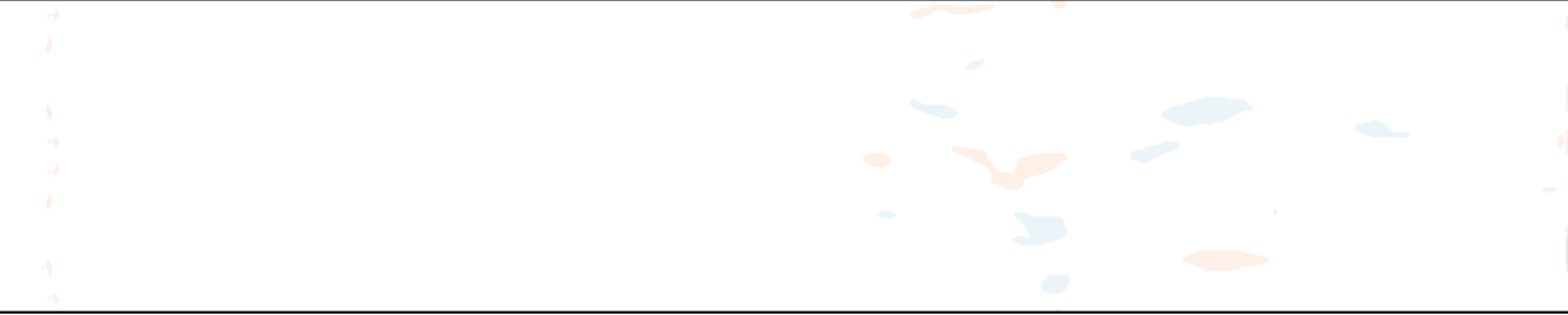}} \\  
 \frame{ \includegraphics[width=0.35\textwidth]{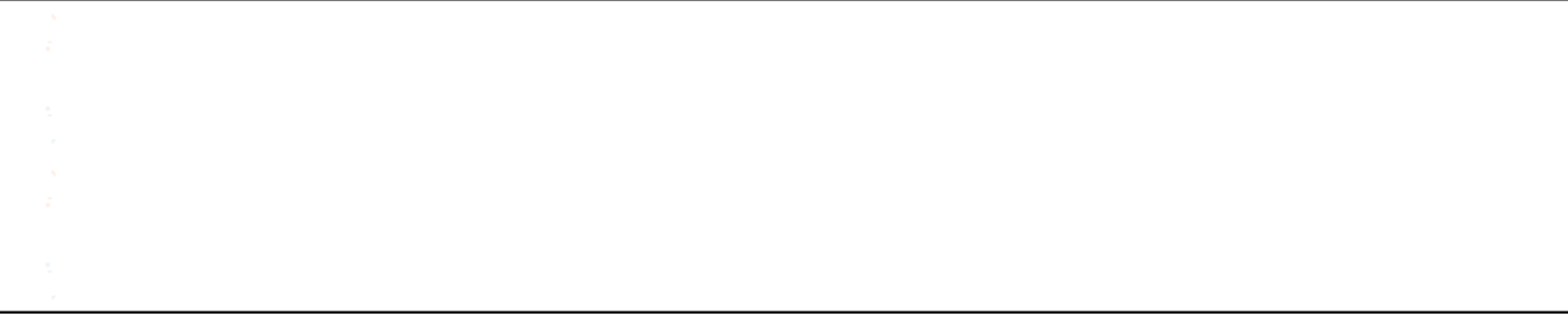}} \\  
 \frame{\includegraphics[width=0.35\textwidth]{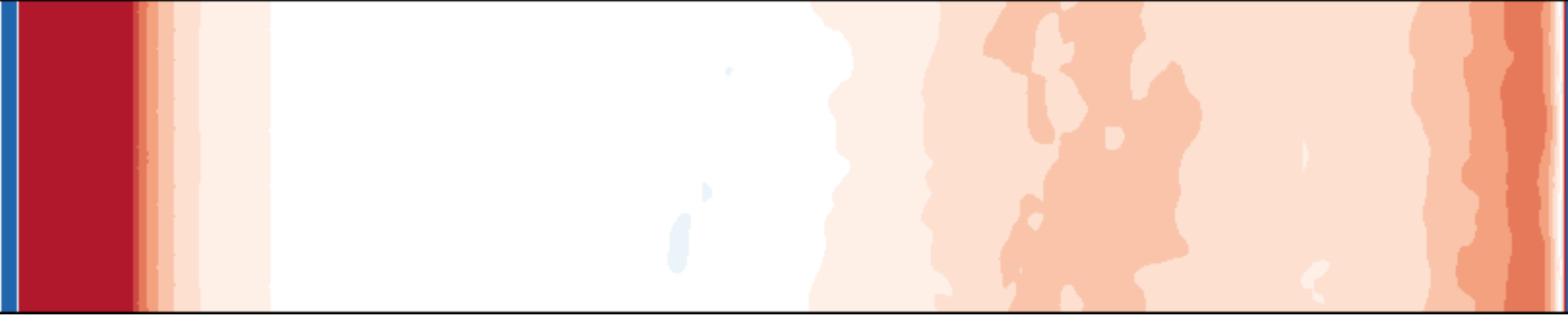}}\vspace{0.02  in}}  \\ \hline
\end{tabular}
\begin{tabular}{ c c c}
 \raisebox{0.25in}{$C_P$} \includegraphics[width=0.275\textwidth]{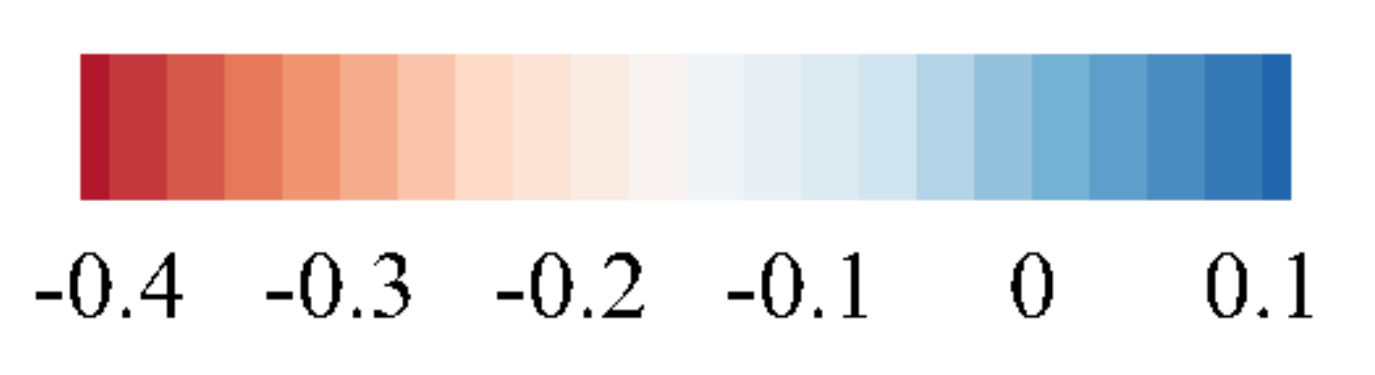} &  
 \raisebox{0.25in}{$\tau_{xy}$} \includegraphics[width=0.275\textwidth]{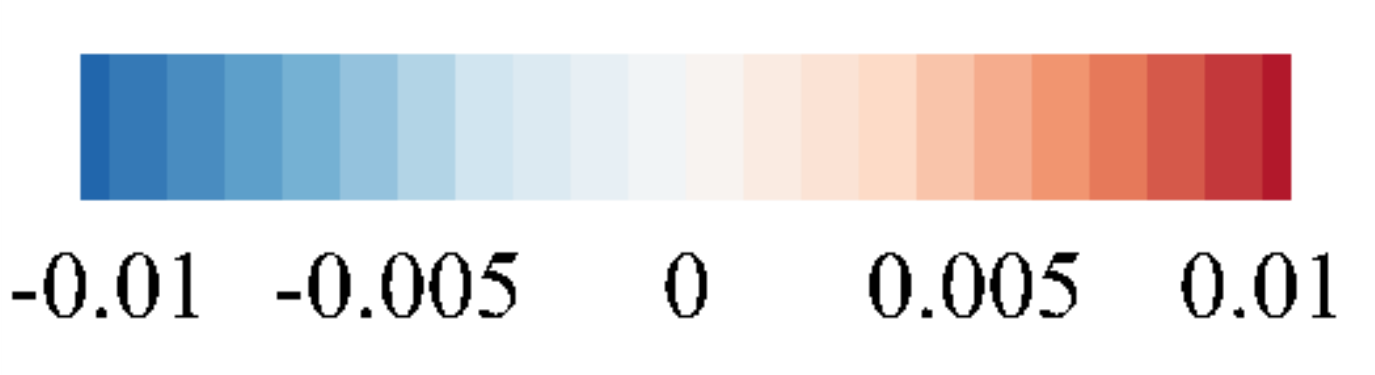} &
 \raisebox{0.25in}{$\sigma_{i}$} \includegraphics[width=0.275\textwidth]{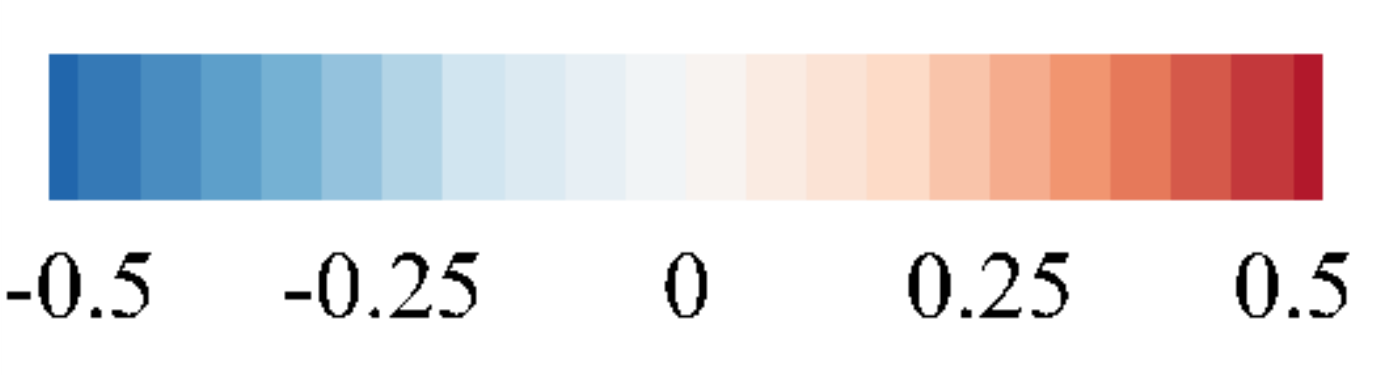}\\ 
\end{tabular} 
}
\caption{Baseline flow. The instantaneous flow field (top left) is visualized with the $Q$-criterion isosurface colored with pressure. The time-averaged zero-streamwise velocity ($\overline{u}_x=0$) iso-surface is colored with spanwise Reynolds stress $\tau_{xy}$ (bottom left).  Visualized on the right are the wall-normal vorticity flux components in the streamwise (top right), wall-normal (middle right), and spanwise (bottom right) directions.}
\label{fig:baseline}
\end{center}
\end{figure}

Let us first examine the baseline flow (without control) over the NACA 0012 airfoil at $\alpha = 9^\circ$ and $Re = 23,000$. Flow over the airfoil at this angle of attack separates near the leading edge and does not reattach over the chord as shown in \fig \ref{fig:baseline} (top left).  Here, we visualize the isosurface of the $Q$-criterion [\citen{Hunt:CTR88}] colored by the pressure coefficient. A shear layer detaches right behind the leading edge and forms large spanwise vortices. The breakdown of these vortices leads to turbulent flow past the mid-chord of the airfoil.  We observe that the low pressure regions above the wing coincides with the cores of large vortical structures.  Towards the trailing edge, bluff body shedding is observed, which is noted by the formation of large-scale, opposite sign vortices. The overall behavior of the separated flow is consistent with the observations by Chang [\citen{Chang76}], Kotapati et al. [\citen{Kotapati:JFM10}] and Kojima et al.~[\citen{Kojima:JA2013}]. 

\begin{figure} 
\begin{center}
\includegraphics[width=0.47\textwidth]{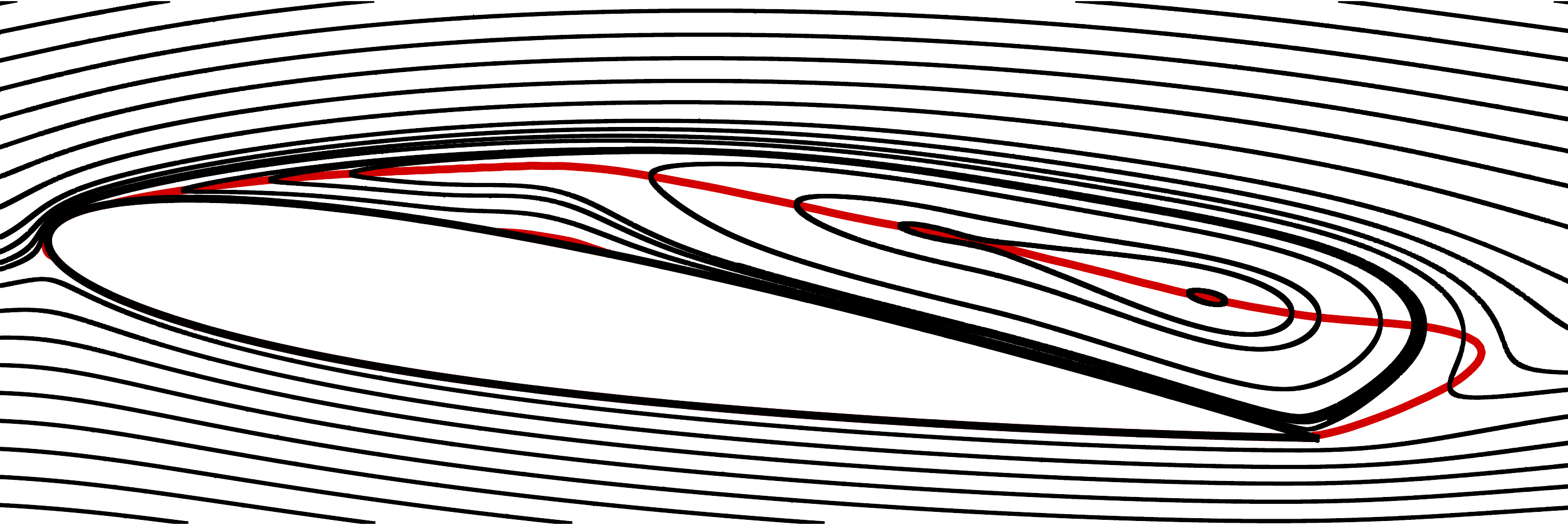}
\caption{Time-averaged streamlines (black) and the time-average, zero-streamwise-velocity ($\overline{u}_x=0 $) curve (red) for baseline flow.}
\label{fig:ZVIS}
\end{center}
\end{figure}

As shown in \fig \ref{fig:baseline} (bottom left), the time-averaged flow reveals that the $\overline{u}_x = 0$ iso-surface covers the entire suction surface of the airfoil, even extends downstream into the wake.  This massive separation over the wing is detrimental to the aerodynamic forces.  The $\overline{u}_x = 0$ iso-surface encompasses the reverse flow region and is used here to capture separation and reattachment.  Note that $\overline{u}_x=0$ iso-surface passes through the center of the time-averaged recirculation zone as illustrated in \fig \ref{fig:ZVIS} and should not be considered as the entirety of a recirculation region.   We also superpose the spanwise Reynolds stress  $\tau_{xy} \equiv \overline{u'_x u'_y}$ (based on the velocity fluctuations, $u_x'$ and $u_y'$, with respect to the mean flow in the $x$ and $y$-directions, respectively) distribution on the $\overline{u}_x=0$ iso-surface in \fig \ref{fig:baseline} (bottom left) to provide insights into turbulent mixing of the fluid between the separated region and the freestream.  For the baseline flow, the time-averaged results are homogeneous in the spanwise direction.

To gain insights into the behavior of separated flow in the near-wall region, we analyze the vortical flux at the surface. For incompressible flow, vorticity can only be generated along the wall surface (or injected by the control inputs).  In specific terms, vorticity is introduced to the flow through a wall-normal diffusive vorticity flux at the surface
\begin{equation}
	\boldsymbol{\Sigma} \cdot \hat{\boldsymbol n}  
	= - \nu (\nabla\boldsymbol{\omega})_0  \cdot \hat{\boldsymbol n}  
	= -\hat{\boldsymbol n} \times \left[ \frac{{\rm d} \boldsymbol{u}_0}{{\rm d}t} + \frac{1}{\rho} (\nabla p)_0 \right] \label{eq:vortFlux},
\end{equation}
which is caused by the acceleration of the wall and the local pressure gradient in the tangential direction [\citen{MortonGAFD84, HornungAFM89, Wu2006}]. Here, the subscript zero denotes the surface value and $\hat{\boldsymbol n}$ represents the wall-normal unit vector. Since the wing is not under acceleration in the present study, only the pressure gradient term contributes to the above flux. This wall-normal diffusive flux is referred to as the source of vorticity [\citen{MortonGAFD84,HornungAFM89}] and is denoted by $\boldsymbol{\Sigma} \cdot\hat{\boldsymbol{n}} \equiv (\sigma_{x}, \sigma_{y}, \sigma_{z})^T$,
where the subscripts of $\sigma_i$ refer to the directions of the wall-normal flux components. Throughout the study, $\sigma_i$ are reported as the time-averaged values to examine the influx and efflux of vorticity from the airfoil surface in the separated region of the flow.

The vorticity flux for the baseline case is shown in \fig \ref{fig:baseline} (right). Since the time-averaged vorticity fluxes are presented here, the streamwise and wall-normal vorticity fluxes ($\sigma_x$ and $\sigma_y$) are negligent for the baseline case. Negative spanwise vorticity flux ($\sigma_z$) is observed upstream of the separation location as the boundary layer negotiates the leading edge. Flow around the leading edge causes a pressure gradient in the streamwise direction, which is correlated to regions with increased spanwise vorticity flux. In what follows, we focus on the separation and reattachment locations (if present), as well as the regions in the vicinity of actuators to examine the effects of control on the modifications of surface vorticity fluxes.

\subsection{Controlled flow}
\label{sec:results9}

\begin{figure}
\begin{center}
{\small
\begin{tabular}{m{0.125\textwidth}m{0.35\textwidth}m{0.35\textwidth}m{0.02\textwidth}} \hline 
\center{ \parbox{0.125\textwidth}{Case \\ $C_\mu$ \\ $u_{\theta,\max}/U_\infty$}} & 
\center{ \parbox{0.35\textwidth}{ {Instantaneous ($Q$, $C_p$)} \\ {Time-averaged ($\overline{u}_x=0$, $\tau_{xy}$)}}} & 
\center{ Surface vorticity flux }  \parbox{0.01\textwidth}{ {$\sigma_{x}$}  \\ $\sigma_{y}$ \\ {$\sigma_{z}$}} &\\  \hline
 \parbox{0.08\textwidth}{A \\ $1.0\%$ \\ $0$}	& 
 \parbox{0.325\textwidth}{\includegraphics[width=0.325\textwidth]{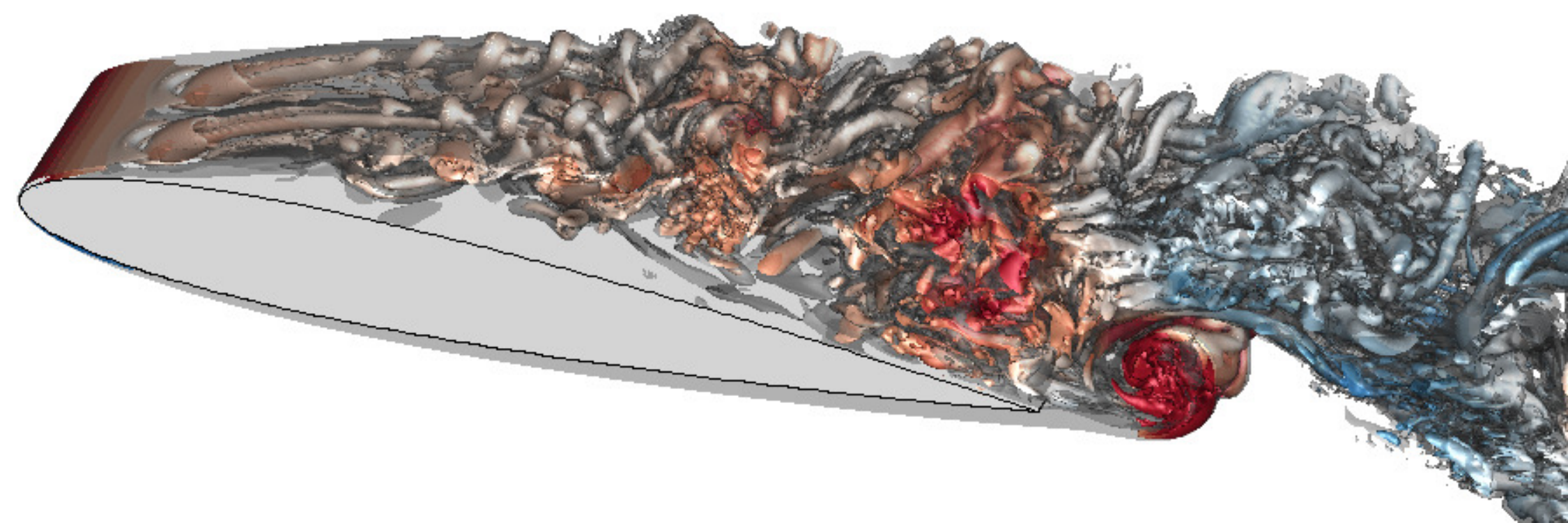} \\ 
 \includegraphics[width=0.325\textwidth]{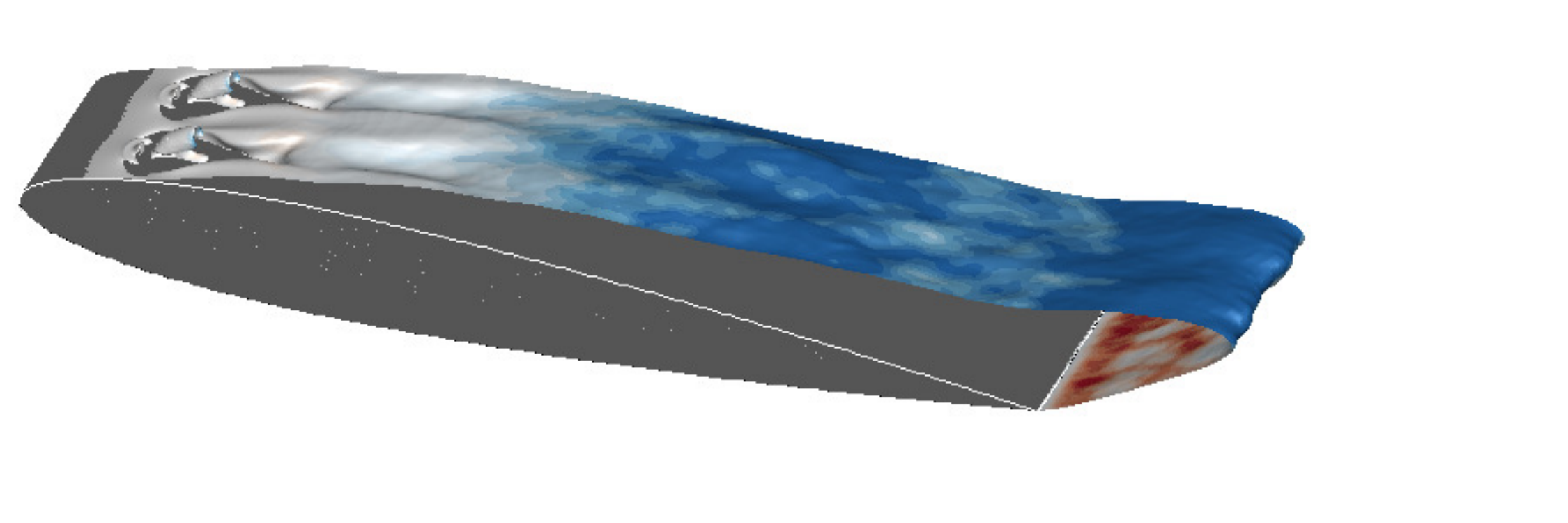}} 	& 
 \vspace{0.02  in} \parbox{0.08\textwidth}{\frame{\includegraphics[width=0.35\textwidth]{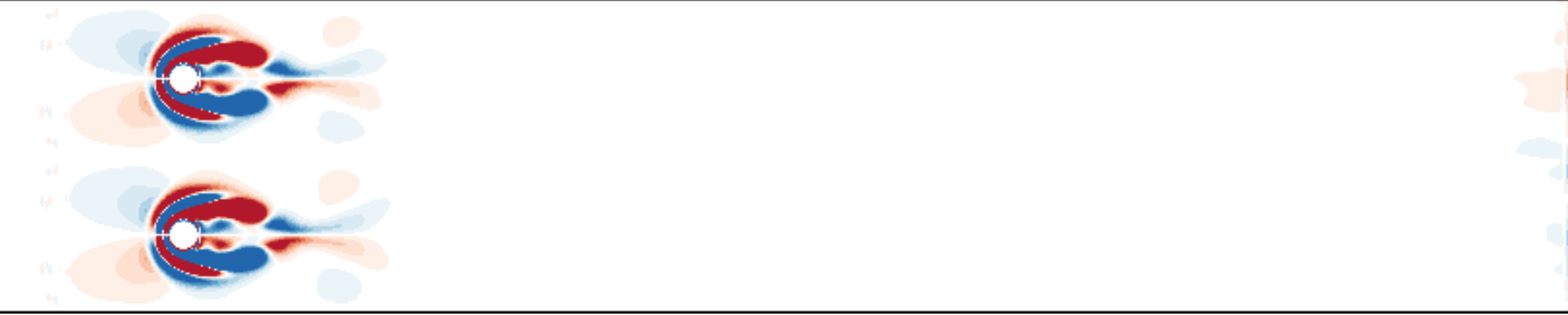}}\\ 
 \frame{\includegraphics[width=0.35\textwidth]{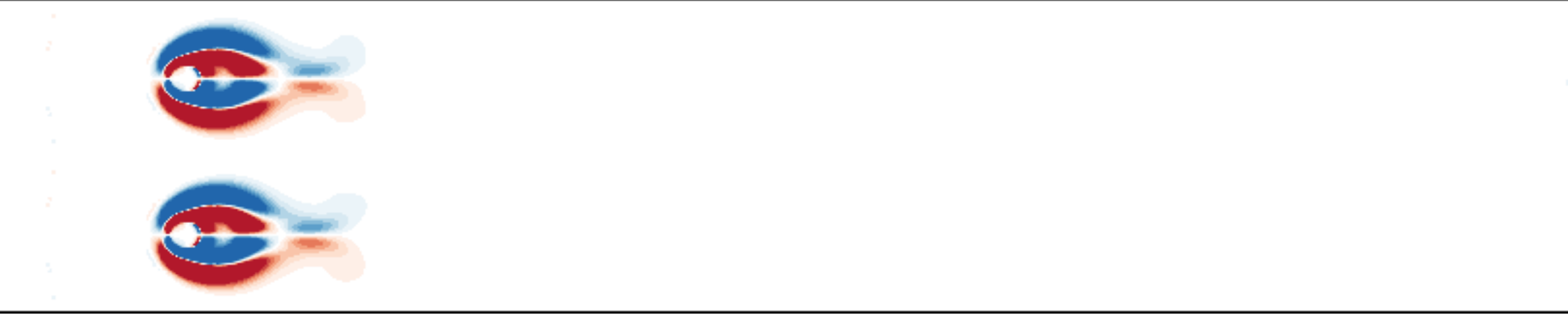}} \\  
 \frame{\includegraphics[width=0.35\textwidth]{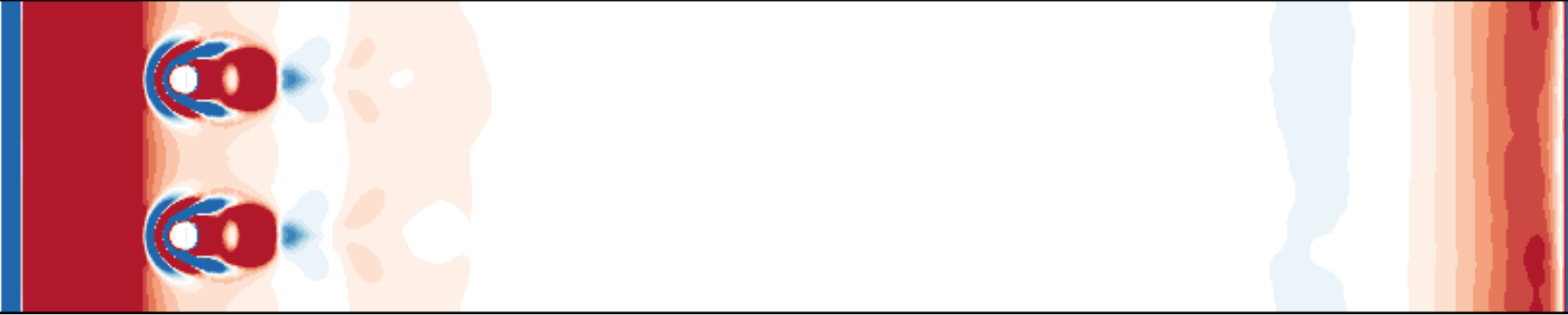}}} \\ \hline 
        \parbox{0.08\textwidth}{B \\ $2.1\%$ \\ $0$}	& 
        \parbox{0.35\textwidth}{\includegraphics[width=0.325\textwidth]{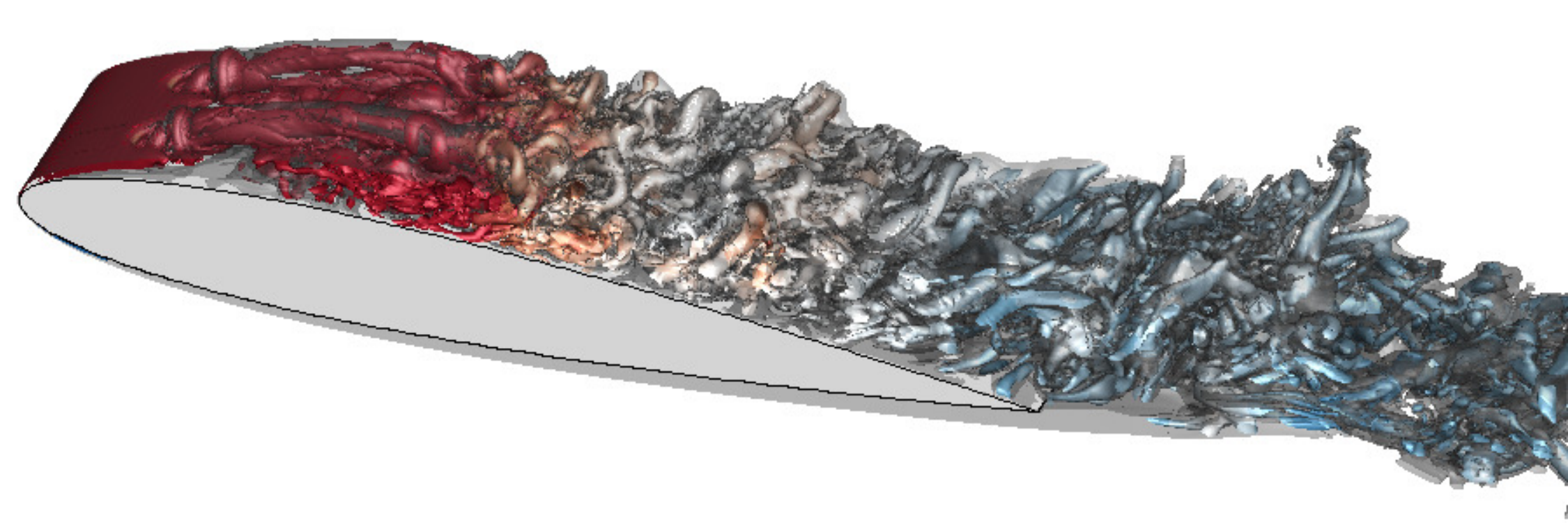}\\ 
        \includegraphics[width=0.325\textwidth]{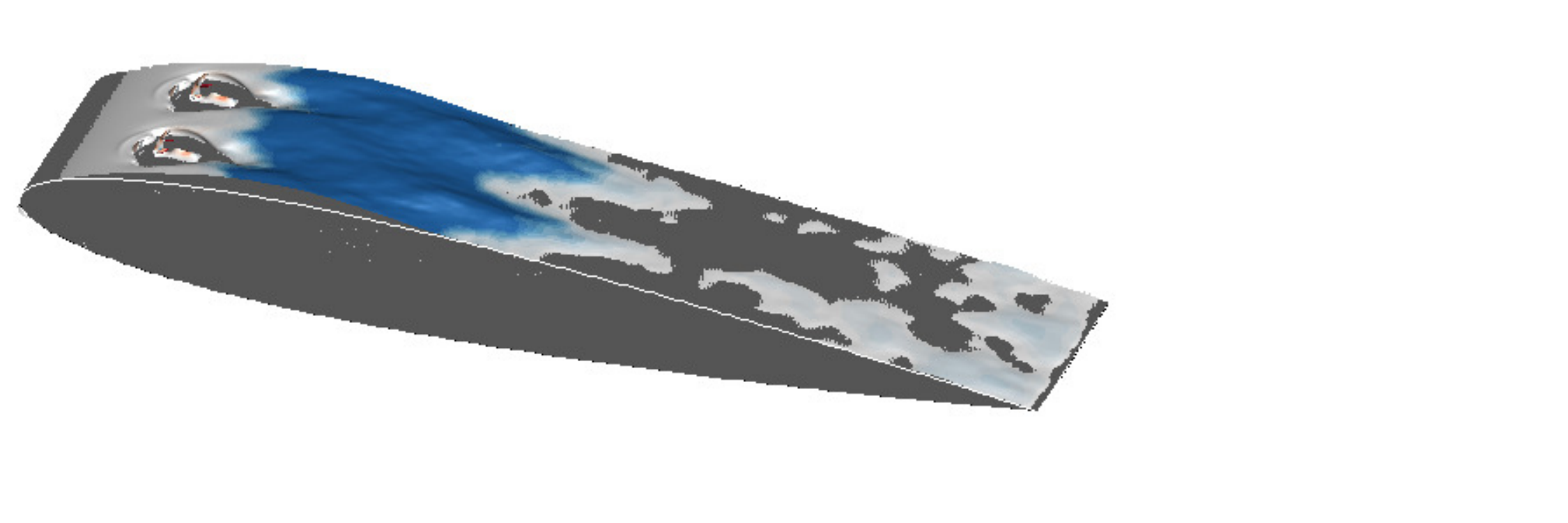}}	& %
        \vspace{0.02  in} \parbox{0.08\textwidth}{\frame{\includegraphics[width=0.35\textwidth]{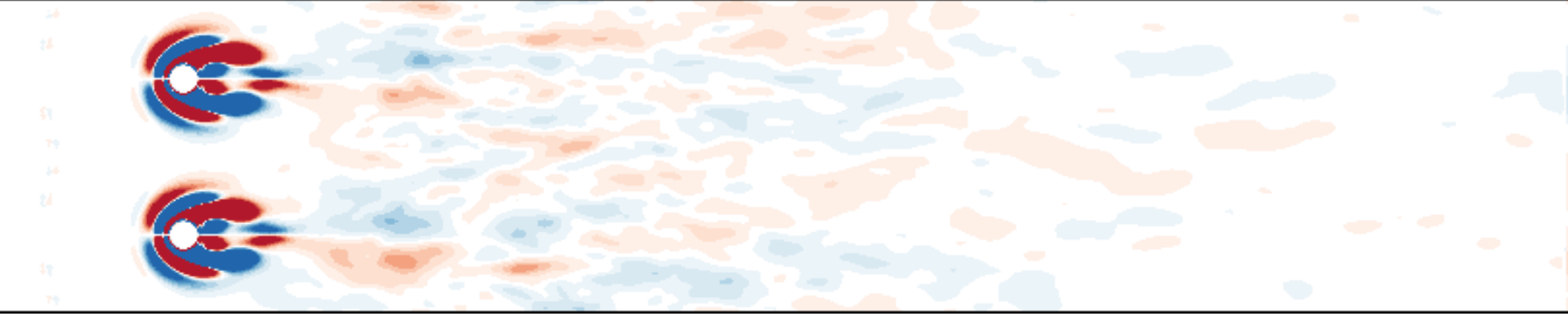}}\\   %
        \frame{\includegraphics[width=0.35\textwidth]{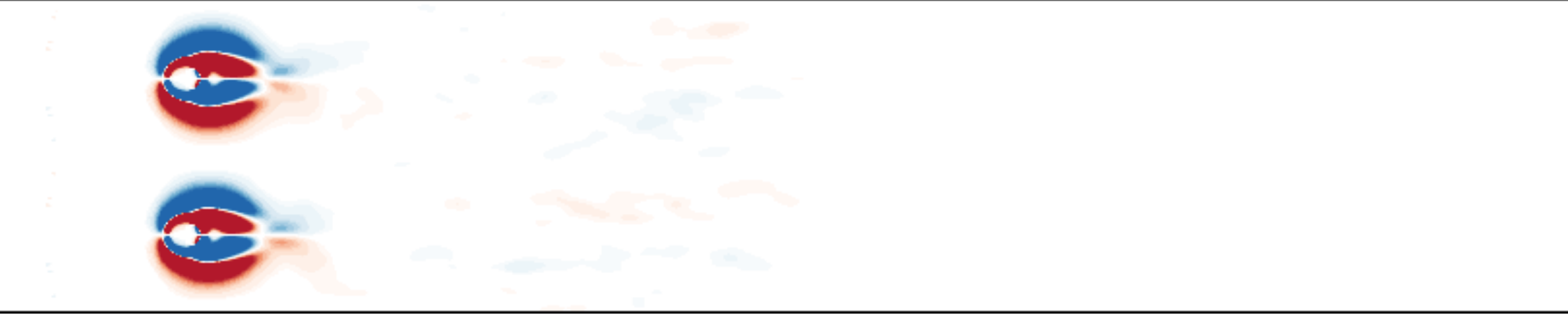}} \\  %
        \frame{\includegraphics[width=0.35\textwidth]{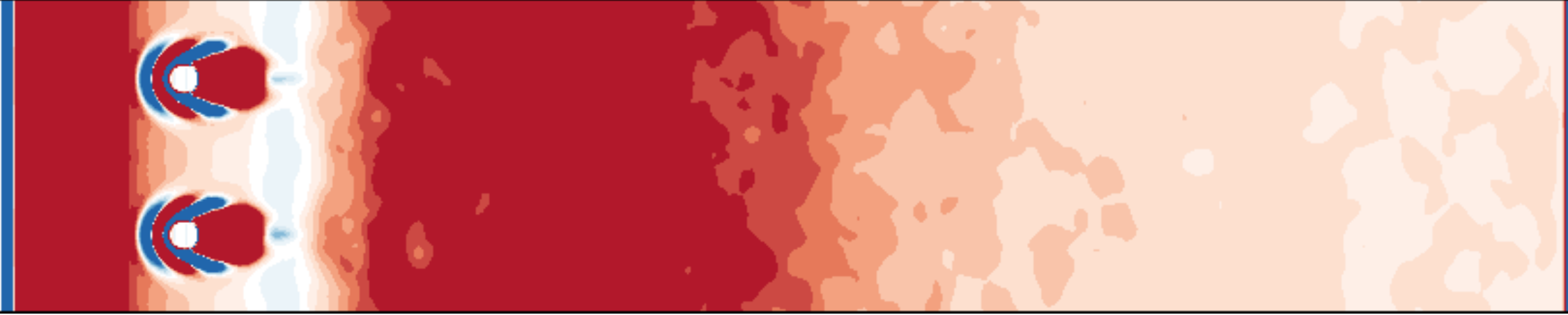}}} \\\hline%
  \parbox{0.08\textwidth}{C \\ $1.0\%$ \\ $0.95$}	& 
  \parbox{0.325\textwidth}{\includegraphics[width=0.325\textwidth]{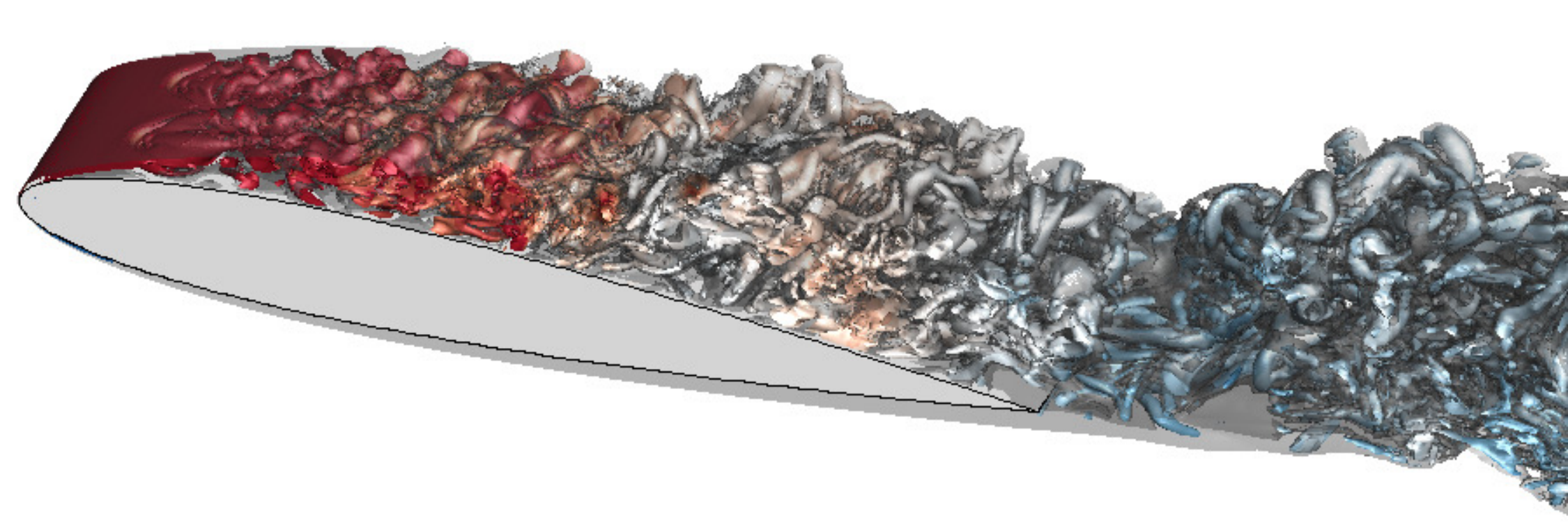} \\   %
  \includegraphics[width=0.325\textwidth]{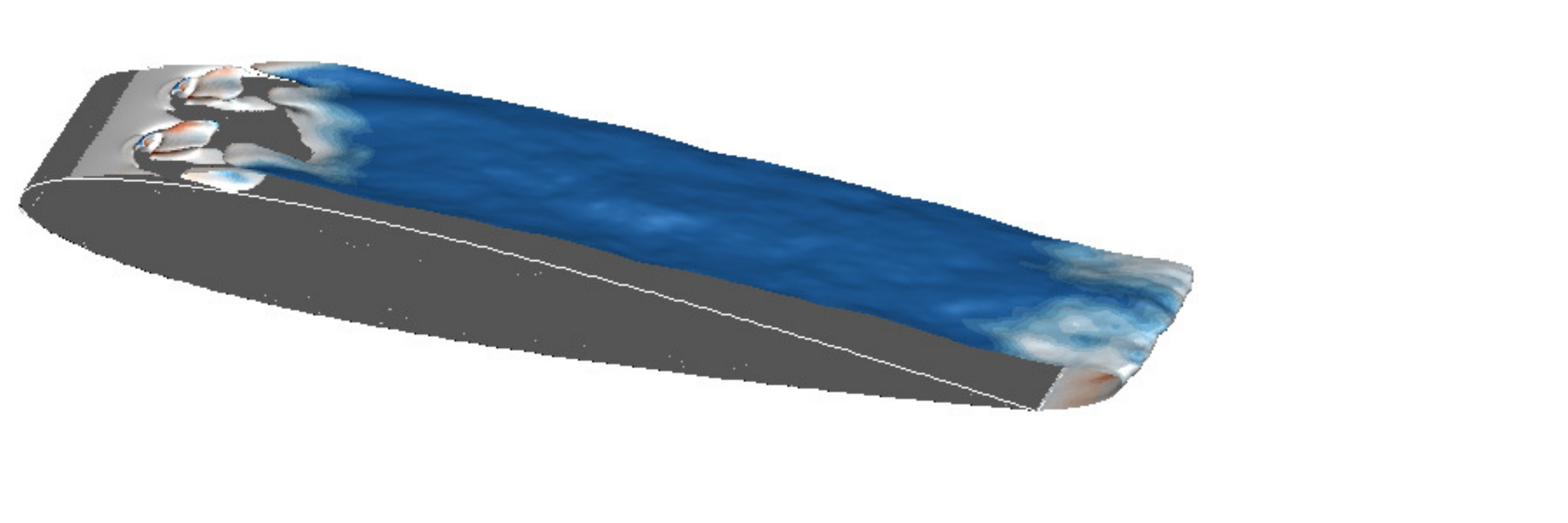}} 		 &%
  \vspace{0.02  in} \parbox{0.08\textwidth}{\frame{\includegraphics[width=0.35\textwidth]{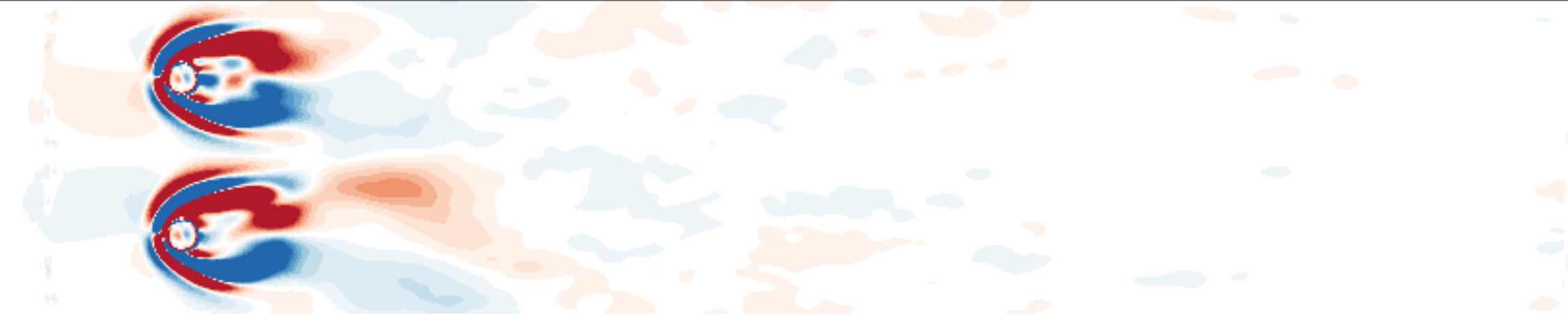}}\\   %
  \frame{\includegraphics[width=0.35\textwidth]{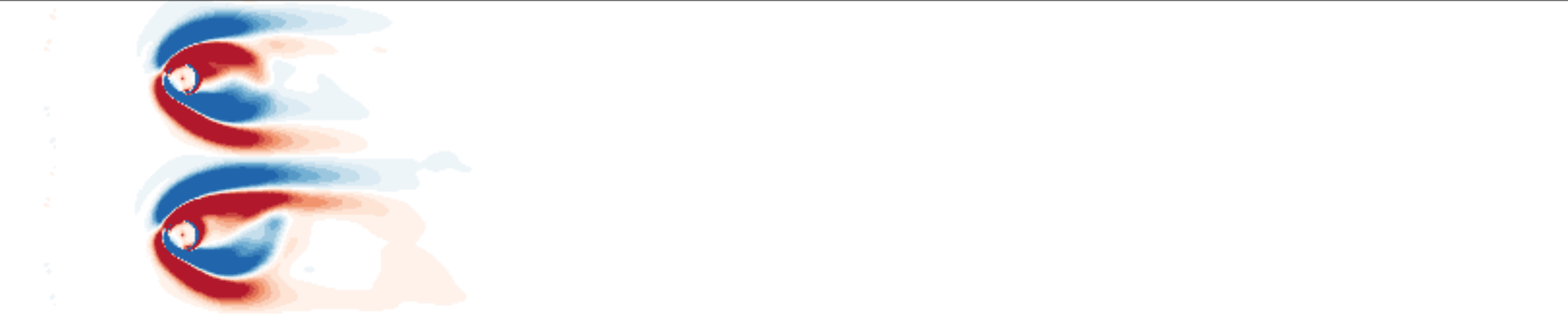}} \\  %
  \frame{\includegraphics[width=0.35\textwidth]{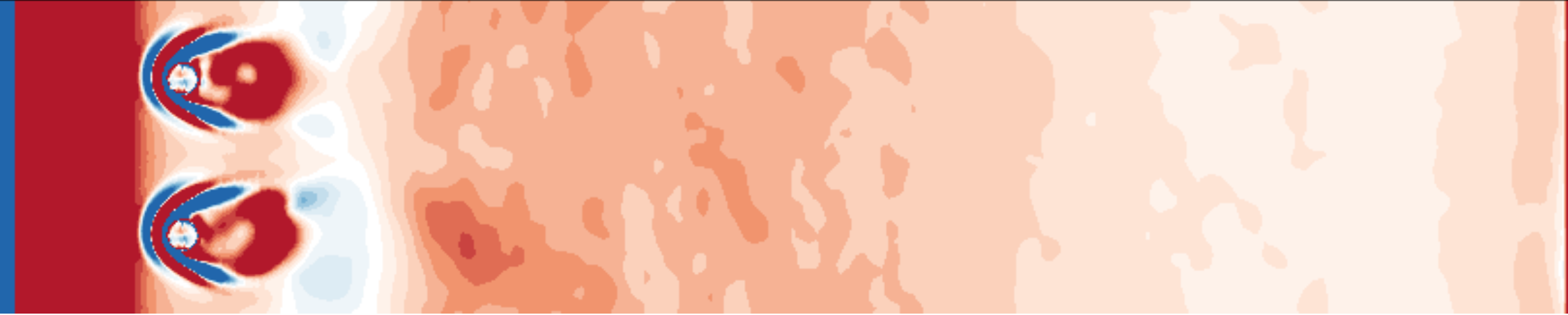}}}  \\\hline%
       \parbox{0.08\textwidth}{D \\ $1.0\%$ \\ $1.26$}	& 
       \parbox{0.35\textwidth}{\includegraphics[width=0.325\textwidth]{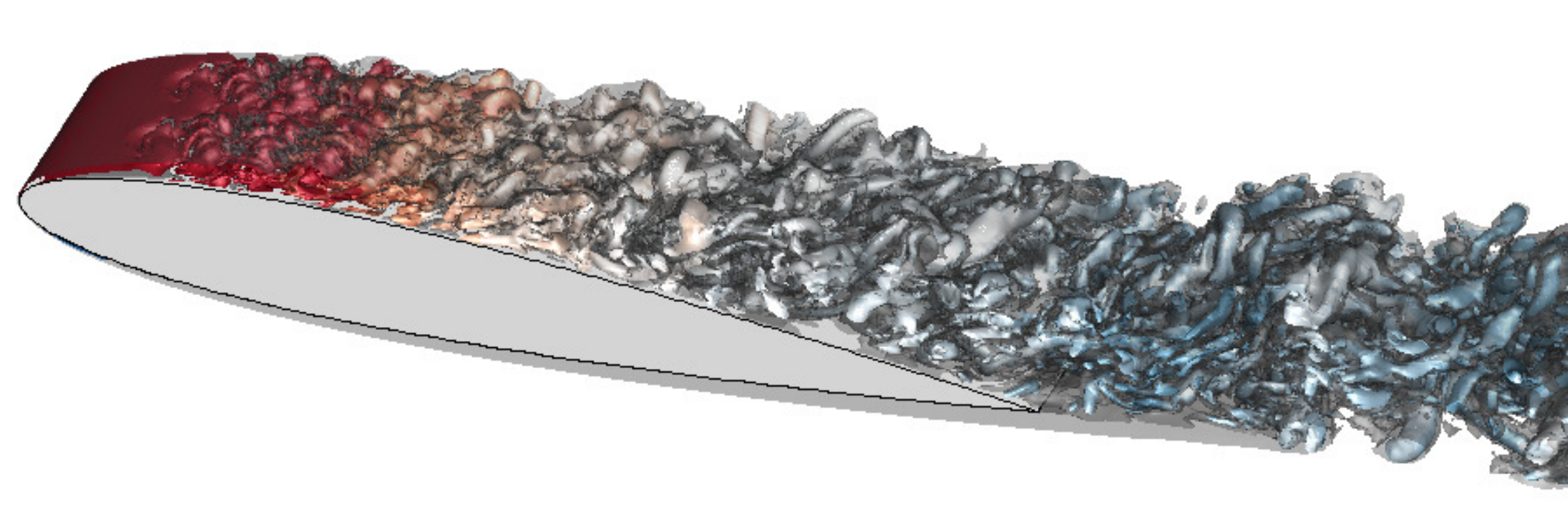} \\ 
       \includegraphics[width=0.325\textwidth]{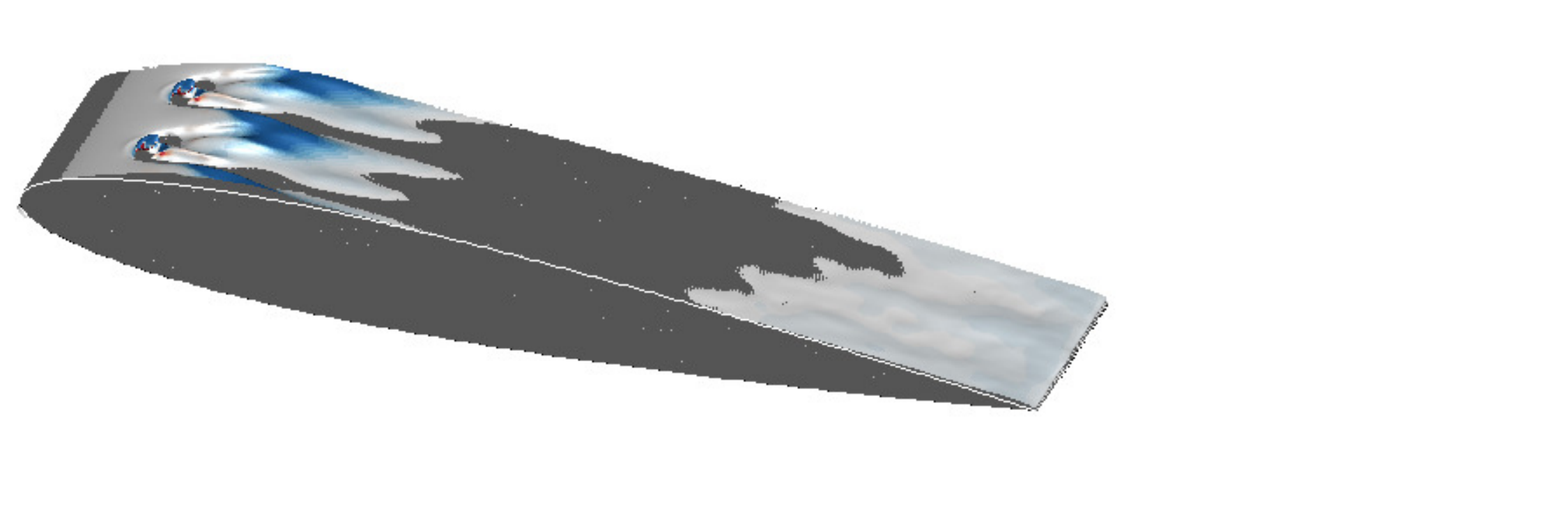}}	&
       \vspace{0.02  in} \parbox{0.08\textwidth}{\frame{\includegraphics[width=0.35\textwidth]{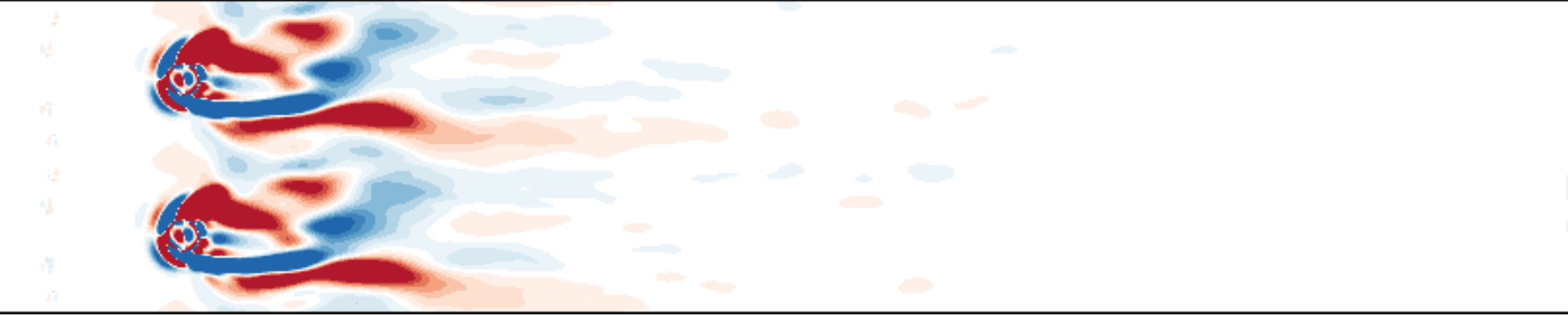}}\\ 
       \frame{\includegraphics[width=0.35\textwidth]{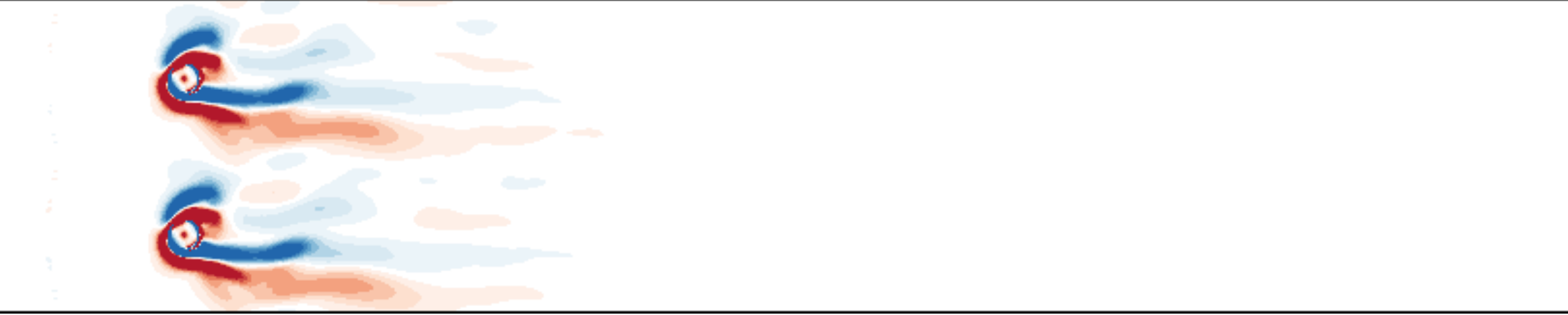}} \\ 
       \frame{\includegraphics[width=0.35\textwidth]{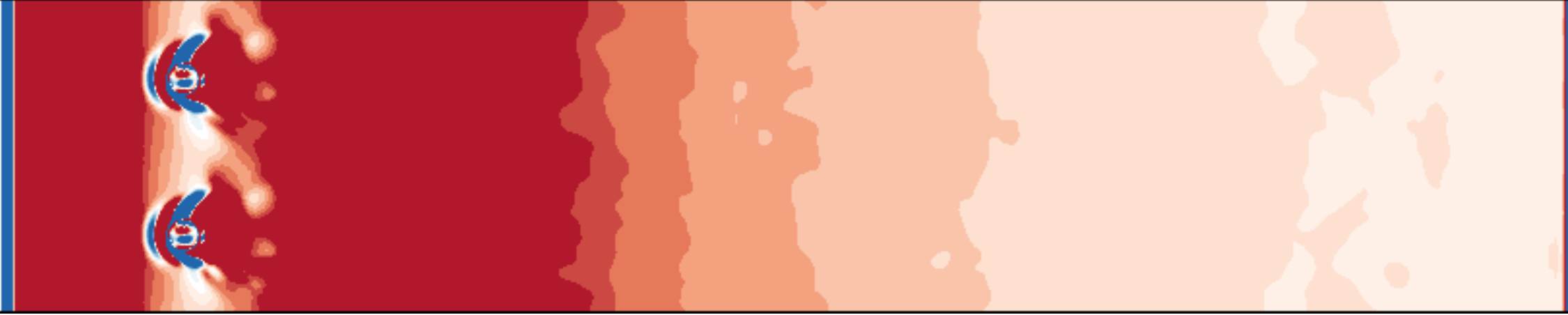}}} \\  \hline \vspace{0.1in} 
\end{tabular}
\begin{tabular}{ c c c}
 \raisebox{0.25in}{$C_P$} \includegraphics[width=0.275\textwidth]{figures/label1} &  
 \raisebox{0.25in}{$\tau_{xy}$} \includegraphics[width=0.275\textwidth]{figures/label2}&  
 \raisebox{0.25in}{$\sigma_{i}$} \includegraphics[width=0.275\textwidth]{figures/label3}\\
\end{tabular}
}
\caption{Controlled flow over a NACA 0012 airfoil at $\alpha = 9^\circ$ and $Re = 23,000$.  Visualizations follow \fig \ref{fig:baseline} for cases A, B, C, and D. See Table \ref{tbl:A9_phi} for full control parameter setups.}
\label{fig:A9_sepReyZ}
\end{center}
\end{figure}

In an effort to reattach the massively separated flow over the airfoil, we introduce steady actuation with maximum wall-normal velocity $u_{n,\max}/U_\infty$ between $0$ and $1.83$ and maximum azimuthal (swirl) velocity values $u_{\theta,\max}/U_\infty$ between $0$ and $2.52$. We list in Table \ref{tbl:A9_phi} four representative control cases that are discussed extensively in this section based on the plots in \fig \ref{fig:A9_sepReyZ}. The entire list of cases considered in this study is provided in Table \ref{tbl:A9_Cases} from Appendix \ref{app:A9}.

\begin{table} 
\centering
\begin {tabular}{l c c c c c c c c c c} \hline
Case 	& $\frac{u_{n,\max}}{U_\infty}$ 	 &$\frac{u_{\theta,\max}}{U_\infty}$ & $C_\mu~[\%]$ & $S$ & $C_\mu^*~[\%]$  & Rot.~dir.  & $C_D$ & $C_{D,P}$ & $C_{D,F}$ & $C_L$\\  \hline\hline	
Baseline& --		& --			& --   & --    & --    & -- 	 & 0.115 & 0.104 & 0.011 & 0.519 \\ \hline
A  & 1.26			& 0				& 1.00 & 0 & 1.00		& --		& 0.108 & 0.095 & 0.013	& 0.403\\ 
B  & 1.83			& 0			    & 2.10 & 0  & 2.10		& --	 	& 0.062	& 0.048 & 0.015 & 0.673\\
C  & 1.26			& 0.95			& 1.00 & 0.47 & 2.17	& ROT 	 	& 0.096	& 0.085 & 0.011	& 0.516\\ 
D  & 1.26			& 1.26			& 1.00 & 0.63 & 2.65	& ROT	 	& 0.069	& 0.057 & 0.012 & 0.665\\ \hline
\end{tabular}
\caption{Representative flow control cases for $\alpha = 9^\circ$ chosen for flow visualization in Fig.~\ref{fig:A9_sepReyZ}.  See Appendix (Table \ref{tbl:A9_Cases}) for the full set of cases considered in the present study.  Pressure and friction drag components, $C_{D,P}$ and $C_{C,F}$, are shown for reference.  The definition of $C_\mu^*$ appears later in Section IV B.}
\label{tbl:A9_phi}
\end{table}


Let us first consider case A where flow control is applied with pure blowing ($u_{\theta, \max} \equiv 0$) using a standard $C_\mu$ value of $1.0\%$, which amounts to $u_{n,\max}/U_\infty = 1.26$. As shown in \fig \ref{fig:A9_sepReyZ}, we observe that the time-averaged separated flow is not significantly modified by pure blowing here.  Comparing case A to the baseline flow in \fig \ref{fig:baseline}, the only noticeable change in the flow is the slight increase in the size of the separated region. Although a deficit in the reverse flow is created in the vicinity of the actuator, the flow eventually separates across the span. In the rear part of the wake, difference between case A and the baseline flow, shown in \fig \ref{fig:baseline} is not noticeable. The actuator jets trigger the Kelvin--Helmholtz instability around the themselves which interacts with the separated shear layer. This results in the spanwise vortex break down close to the leading edge.  We observe surface vorticity flux in case A modified near the actuators by the change in pressure gradients but not in a significant manner in the separated region.

The sole addition of momentum with a wall-normal velocity of $u_{n,\max}/U_\infty = 1.26$ is not effective in overcoming the adverse pressure gradient and reattaching the flow as shown in case A. Further increasing the wall-normal velocity to $u_{n,\max}/U_\infty = 1.83$ (case B, $C_\mu = 2.1\%$) diminishes the size of the reverse flow region. Case B produces well-formed actuator jets similar to case A, but the increased velocity causes the localized instability to appear closer to the actuator injection ports.  The interaction of this localized instability with the spanwise vortices is desired for the breakdown of the large vortical structures.  This change in the flow produces significant Reynolds stress near the actuators, which is correlated with the mixing of freestream and boundary layer.  Note that the breakup of the spanwise vortices also can be considered as a forced transition process which has similarity to the transition in boundary layers [\citen{Liu:CF14}].  However, the controlled flow cases in the present study do not exhibit $\Lambda$ vortices.  Downstream of each actuator, we observe slightly higher level of streamwise vorticity flux, which is indicative of counter-rotating vortices forming from each actuator jet aiding in the reattachment of the flow. Both lift and drag forces in case B are improved significantly compared to the baseline and case A. 

Next, reverting back to a wall-normal velocity $u_{n,\max}/u_\infty = 1.26$, let us now superpose angular momentum (swirl) to the control jets. As shown in \fig \ref{fig:A9_sepReyZ}, the combination of wall-normal and angular momentum injections (cases C and D) can greatly modify the flow field. Cases C and D use azimuthal actuator velocity profiles with $u_{\theta,\max}/U_\infty = 0.95$ and $1.26$, respectively.  We notice that the larger actuator jet structures seen in the pure blowing case are closer to the wing surface and are broken into smaller structures when rotation is added to the actuation input. For these two cases, increasing azimuthal velocity input decreases the size of the reverse flow region. For case C, flow still remains separated behind the actuators but the resulting size of the separated flow region is noticeably smaller than the baseline flow and case A. The change in the time-averaged flow field for case C is reflected in the forces. Drag is further decreased and lift is increased compared to forcing with pure blowing with identical wall-normal velocity input.

Full reattachment of the flow with the same wall-normal velocity can be achieved using higher azimuthal velocity input. The control input with $u_{n,\max}/U_\infty = 1.26$ and $u_{\theta,\max}/U_\infty = 1.26$ by case D fully reattaches the flow downstream of the actuators. We observe in \fig \ref{fig:A9_sepReyZ} (case D) flow separates over a small region upstream of the actuators, and the momentum added to the boundary layer by the actuators allows for the flow to overcome the adverse pressure gradient. We hence observe a diminished reverse flow region downstream of the actuators. The result of the attached flow translates to significant improvements in terms of aerodynamic forces. It should be observed that $\sigma_{z}$ contour plot for case D shows a desirable profile over the entire top surface, indicating favorable pressure gradient (attached flow) achieved by control inputs with wall-normal momentum and angular momentum inputs. 

Wall-normal injections in Case A perturb the shear layer but does not significantly impact the separated wake. The addition of angular momentum reduces the size of the separated flow region and, with sufficient level, it can completely reattach the flow.  From case C with $u_{\theta,\max}/U_\infty = 0.95$, we observe a decrease in the size of the separated flow region. Moreover, in case D with $u_{\theta,\max}/U_\infty = 1.26$, separation is effectively eliminated, yielding $37\%$ drag decrease and $31\%$ lift increase. The same result of attached flow is attainable for a larger wall-normal velocity value $u_{n,\max}/U_\infty = 1.83$, seen in case B.  Below, we consider how the two independent input velocity magnitudes can be consolidated into a single non-dimensional forcing parameter to characterize the effectiveness of the flow control inputs.

Let us briefly discuss the influence of control on drag.  In Table \ref{tbl:A9_phi}, we present the pressure and friction drag components ($C_{D,P}$ and $C_{D,F}$, respectively) for Cases A to D.  From the baseline and all controlled cases at this angle of attack, drag is comprised primarily of pressure drag.  The addition of control reduces the pressure drag for all cases but the friction drag stays fairly constant across all cases.  For the cases in which reattached flow is achieved (Cases B and D), pressure drag is drastically decreased due to removal of flow separation.


\section{Control input quantification}
\label{sec:Quantification}

In the previous section, we have discussed that the combined use of wall-normal and angular momentum injections can effectively suppress flow separation over a NACA0012 airfoil.  Here, we consider a much larger number of flow control cases than those presented above and present an approach to quantify the two independent control inputs of wall-normal momentum and angular momentum through a single consolidated control input parameter.  In what follows, we extend the standard definition of the coefficient of momentum to incorporate the influence of angular momentum (swirl) injection and show that the behavior of the lift enhancement under flow control can be captured well in an integrated manner.  Moreover, we demonstrate how the considered cases can be classified into three major categories of separated, transitional, and reattached flows based on the modified definition of the coefficient of momentum.  All cases considered in this section are compiled in Table \ref{tbl:A9_Cases} in the Appendix.  

\subsection{Coefficient of momentum}

\begin{figure}
\begin{center}
\includegraphics[width=0.5\textwidth]{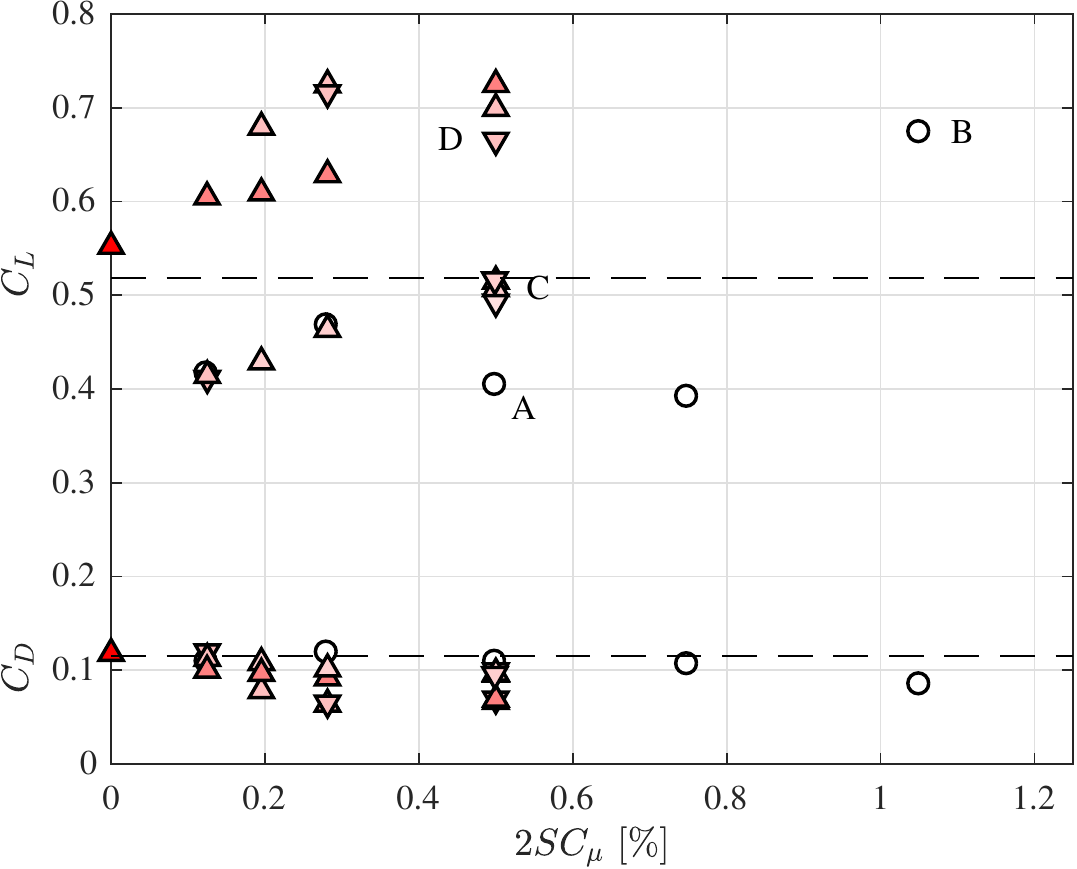} 
\raisebox{0.25 in}{
\includegraphics[trim={8.5cm 0 6cm 0},clip,width=0.125\textwidth]{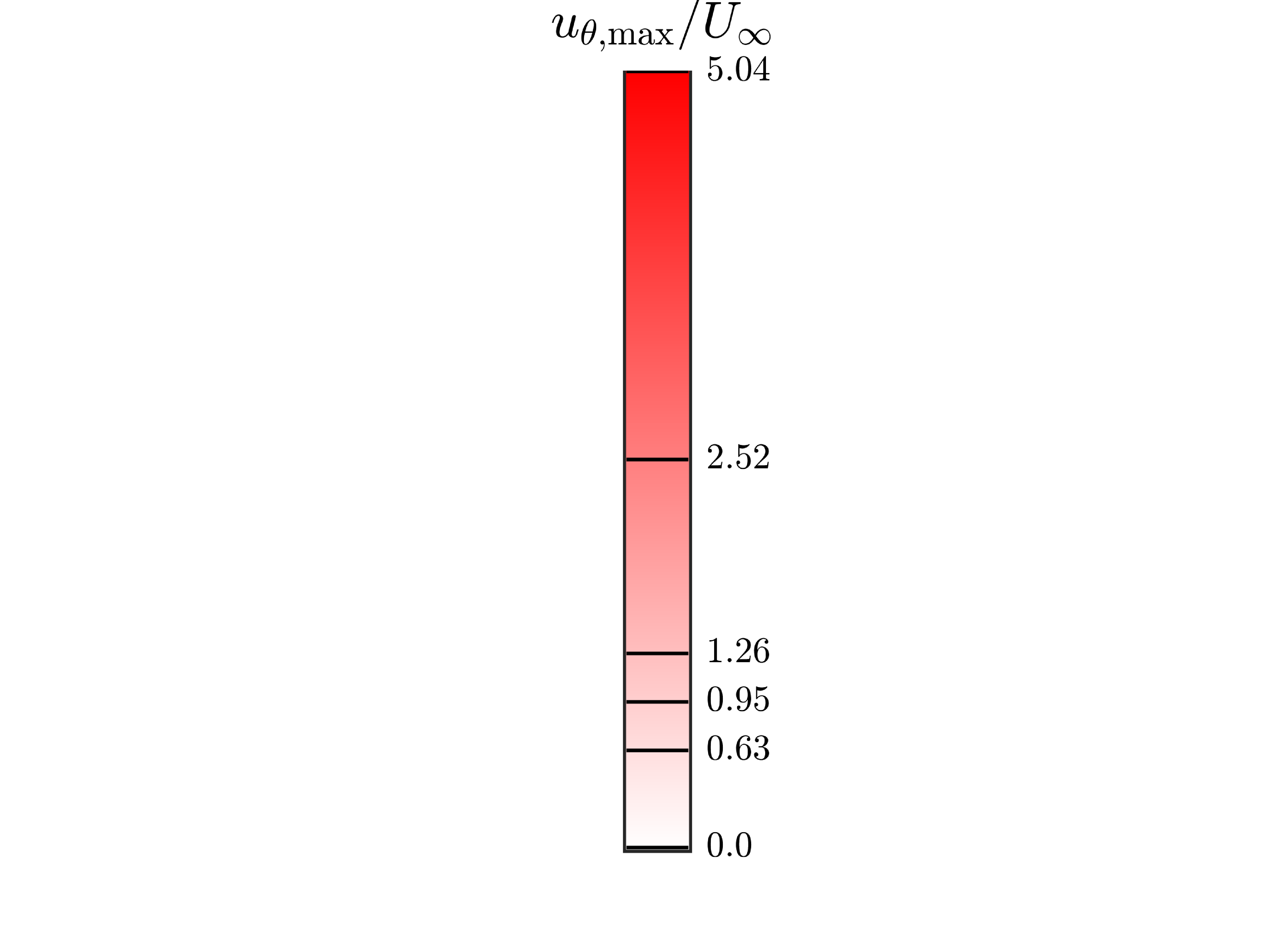}} 
		\caption{The coefficients of drag and lift versus the coefficient of momentum. The baseline values are indicated by \dashed ~and the controlled cases include pure blowing ($\bigcirc$), co-rotating ($\triangledown$), and counter-rotating ($\triangle$).}
		\label{fig:Force_CM}
\end{center}
\end{figure}

The coefficient of momentum $C_\mu$, defined earlier in \eq (\ref{eq:Cmu}), is arguably the most commonly used non-dimensional parameter to quantify the control input in active flow control. Let us start by examining all considered cases with respect to the coefficient of momentum, $C_\mu$.  The modifications of lift and drag forces with flow control for different levels of wall-normal momentum input are summarized in \fig \ref{fig:Force_CM} over the range of $0\% \le C_\mu \le 2.1\%$.  This range is selected to follow the previous studies achieving effective flow control over symmetric airfoils [\citen{Deng:CF07,Gilarranz:ASME05, Seifert:AIAA99,You:PF08}]. 

Shown by the color of the symbols in this figure is the magnitude of angular velocity input $u_{\theta,\max}/U_\infty$.  As discussed above, we expect the change in the angular velocity input to provide variations in the lift enhancement achieved for a fixed $C_\mu$ input. We also indicate the arrangements of the swirling direction with $\bigcirc$ (pure blowing), $\triangledown$ (co-rotating), and $\triangle$ (counter-rotating).  For the same angular momentum input level, we observe that there are only small differences in the lift and drag values between the controlled cases with co-rotating and counter-rotating jets.  For all of the cases examined in this study, the behavior of the drag force shows a decreasing trend with respect to increasing coefficient of momentum.  Although the reduction in drag force may appear minor ($\Delta C_D \approx -0.04$), the percentage change in drag reduction with respect to the baseline value is substantial ($-35\%$).  While the majority of the controlled cases reduced drag on the airfoil, lift enhancement is achieved only with appropriate selection of control inputs.

For $0.4\% \lesssim C_\mu \lesssim 1\%$, the addition of angular momentum influences the resulting flow as it can be seen by the variation in achieved lift enhancement. In general, for a fixed $C_\mu$, use of larger $u_{\theta,\max}/U_\infty$ value results in increased lift and decreased drag. For example, cases A, C, and D all use the same value of momentum coefficient, and we observe that increasing azimuthal velocity input by the actuator increases lift and decreases drag. The enhancements in force values are related to the diminishing size of the reverse flow region. There is saturation in achieved lift increase as seen with cases B and D. Once the flow is reattached, further improving aerodynamic forces is difficult. 


\subsection{Modified coefficient of momentum}
\label{sec:Cswirl}
\label{sec:modCmu}

We have observed above that effectiveness of actuation is dependent on the combination of wall-normal momentum and angular momentum inputs.  The sole use of $C_\mu$ cannot fully describe the overall enhancement in lift force achieved with these control inputs.   As the spread of data over $C_\mu$ in \fig \ref{fig:Force_CM} suggests, the influence of swirling input also needs to be accounted for to quantify the lift enhancement.  Here, we derive a modified coefficient of momentum that incorporates both the wall-normal and angular momentum components of the actuation inputs to capture the lift enhancement achieved from flow control in a consolidated manner.

Let us consider a modified characteristic velocity $u_j^*$ that takes into consideration the influence of added swirl (angular momentum) for each actuator.  We relate this modified characteristic velocity $u_j^*$ to the wall-normal jet velocity $u_{n,\max}$ and the swirl velocity $u_{\theta,\max}$.  That is, we represent the modified characteristic jet velocity to be 
\begin{equation}
   u_j^* = u_{n,\max} \left[ 1 + S(u_{\theta,\max}, u_{n,\max}) \right] = u_{n,\max} \left[ 1 + S\left(\frac{u_{\theta,\max}}{u_{n,\max}}\right) \right] ,
   \label{ref:u_characteristic1}
\end{equation}
where $S$ is a correction function that takes the azimuthal momentum input into consideration.  Note that the correction function $S$ must be non-dimensional, which suggests that this function should be related to the non-dimensional ratio of $u_{\theta,\max}/u_{n,\max}$. Use of this non-dimensional ratio ensures that as $S\rightarrow 0$ we recover the characteristic velocity of $u_{n,\max}$.  Moreover, the characteristic velocity should take into account the importance of angular momentum injection relative to the wall-normal momentum input.  Hence, we evaluate $S$ as the non-dimensional swirl number [\citen{Lilley1985, PandaPF94}] defined by 
\begin{equation}
   S \equiv \frac{G_\theta}{r_a G_n},
   \label{eq:swirl_def}
\end{equation} 
which quantifies the ratio between the wall-normal flux of tangential momentum $G_\theta$ and the wall-normal flux of wall-normal momentum $G_n$, evaluated as
\begin{equation}
   G_\theta = 2 \pi \rho_\infty \int_{r=0}^{r_a} u_n u_\theta r^2 {\rm d}r,
   \quad
   G_n = 2 \pi \rho_\infty \int_{r=0}^{r_a} u_n^2 r {\rm d}r,
\end{equation}
respectively.  For the velocity profiles selected in this study, \eq (\ref{eq:vel_profiles}), the non-dimensional swirl number becomes 
\begin{equation}
   S = \kappa \frac{u_{\theta,\max}}{u_{n,\max}}, 
   \quad \text{where}
   \quad \kappa = \frac{22}{35}.
   \label{eq:Sdef}
\end{equation} 
As such, we find that the correction function $S$ indeed is represented as a function of the velocity ratio $u_{\theta,\max}/u_{n,\max}$.  It is also possible to express this swirl ratio $S$ in terms of the ratio of average velocity values, which will yield a different scaling parameter for $\kappa$.  The majority of the controlled cases considered in this study are performed with the swirl number of $S<1$.

With the modified characteristic jet velocity $u_j^*$ now derived with the effect of angular momentum incorporated, we let
\begin{equation}
   u_{n,\max} \rightarrow u_j^* = u_{n,\max} (1 + S).
\end{equation}
and evaluate the modified momentum coefficient $C_\mu^*$ using $u_j^*$ in the numerator of its definition
\begin{equation}
   C_\mu^* \equiv \frac{\rho_\infty u_j^{*2} A_j N_\text{act}}{\frac{1}{2}\rho_\infty U_\infty^2 A}
   = \frac{\rho_\infty u_{n,\max}^2 (1+S)^2 A_j N_\text{act}}{\frac{1}{2}\rho_\infty U_\infty^2 A}.
   \label{eq:Cmustar1}   
\end{equation}
This modified coefficient of momentum can then be expressed as
\begin{equation}
   C_\mu^* = (1+S)^2 C_\mu
   = C_\mu + 2S C_\mu + S^2 C_\mu.
   \label{eq:Cmustar2}
\end{equation}
The first term on the right hand side retains the original coefficient of momentum $C_\mu$, while the two additional terms on the right hand side account for the influence of swirl injection.  We should note that this modified coefficient of momentum $C_\mu^*$ extends the original coefficient in hopes of incorporating the role of swirl injection for capturing the combined flow control effects over a reduced space.  Although similar in spirit, one should be aware that this viewpoint differs somewhat from the original objective of $C_\mu$, which quantifies the cost of introducing steady actuation jets.

Using \eq (\ref{eq:Sdef}), the two correctional terms $2S C_\mu$ and $S^2 C_\mu$ in \eq (\ref{eq:Cmustar2}) become
\begin{align}
   & 2S C_\mu = \frac{\rho_\infty (2\kappa u_{n,\max} u_{\theta,\max}) A_j N_\text{act}}{\frac{1}{2}\rho_\infty U_\infty^2 A},
   \label{eq:correct1}
   \\
   & S^2 C_\mu =  \frac{\rho_\infty (\kappa u_{\theta,\max})^2 A_j N_\text{act}}{\frac{1}{2}\rho_\infty U_\infty^2 A},
   \label{eq:correct2}   
\end{align}
respectively.  The term $2S C_\mu$ captures the cross-product relationship between the two actuation inputs and $S^2 C_\mu$ incorporates $u_{\theta,\max}^2$ with a scaling factor $\kappa^2$.  Hence, we emphasize that the modified coefficient of momentum $C_\mu^*$ is not comprised simply by the linear superposition of the square terms $u_{n,\max}^2$ and $u_{\theta,\max}^2$.  The cross term $u_{n,\max} u_{\theta,\max}$ is indeed required to effectively capture the influence of wall-normal and swirling injections, as we shall observe later.

\begin{figure}
\begin{center}
\includegraphics[width=0.5\textwidth]{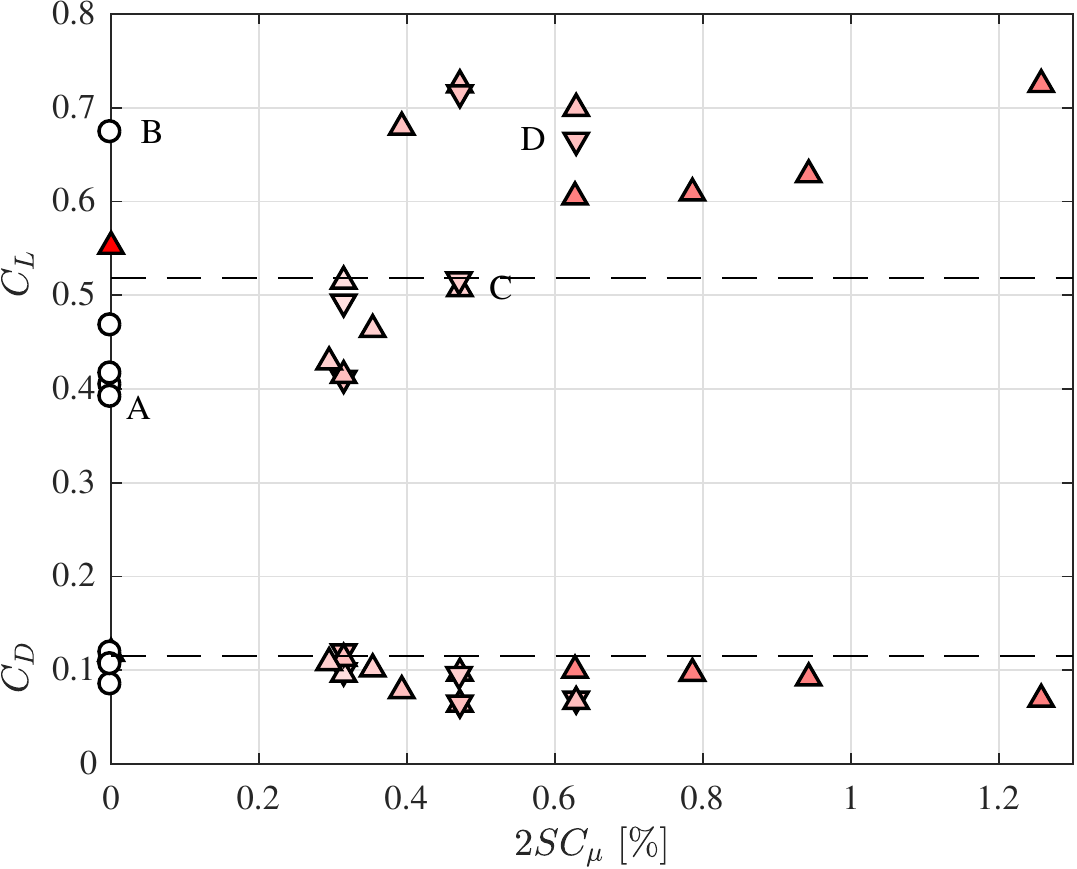} 
\raisebox{0.25 in}{
\includegraphics[trim={8.7cm 0 6cm 0},clip,width=0.125\textwidth]{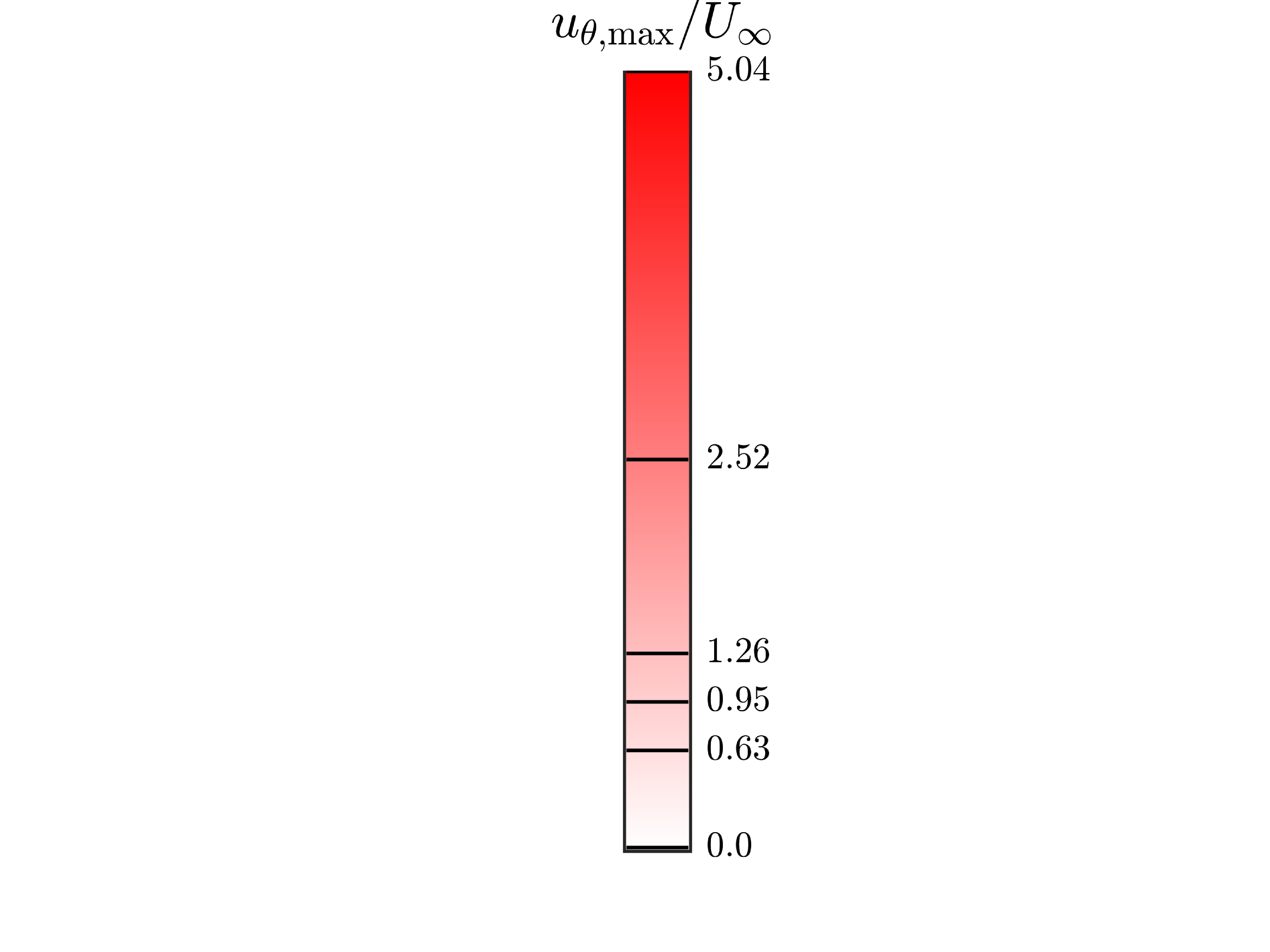}} 
		\caption{The correction term $2S C_\mu$ versus the lift and drag forces.
		 The baseline values are indicated by \dashed~ and the controlled cases include pure blowing ($\bigcirc$), co-rotating ($\triangledown$), and counter-rotating ($\triangle$).}
		\label{fig:CoeffSwirl}
\end{center}
\end{figure}
 
\begin{figure}
\begin{center}
\includegraphics[width=0.5\textwidth]{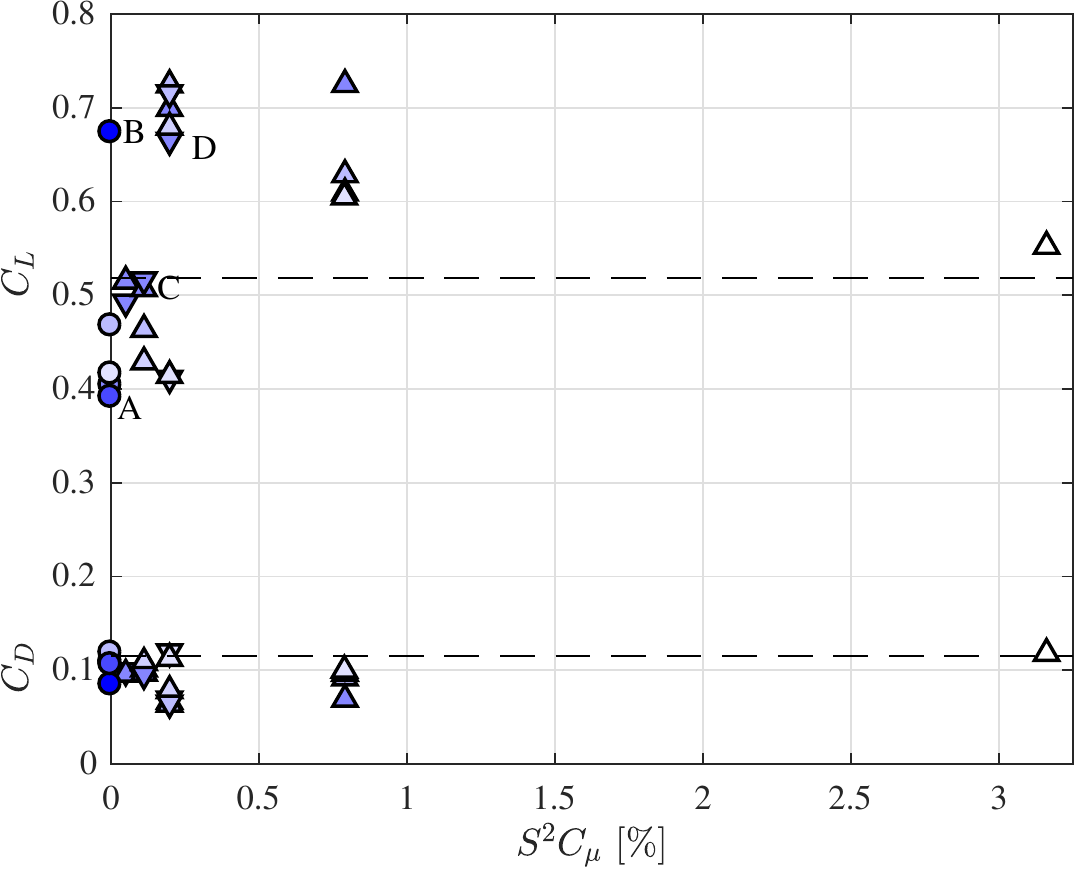} 
\raisebox{0.25 in}{
\includegraphics[trim={8.5cm 0 6cm 0},clip,width=0.125\textwidth]{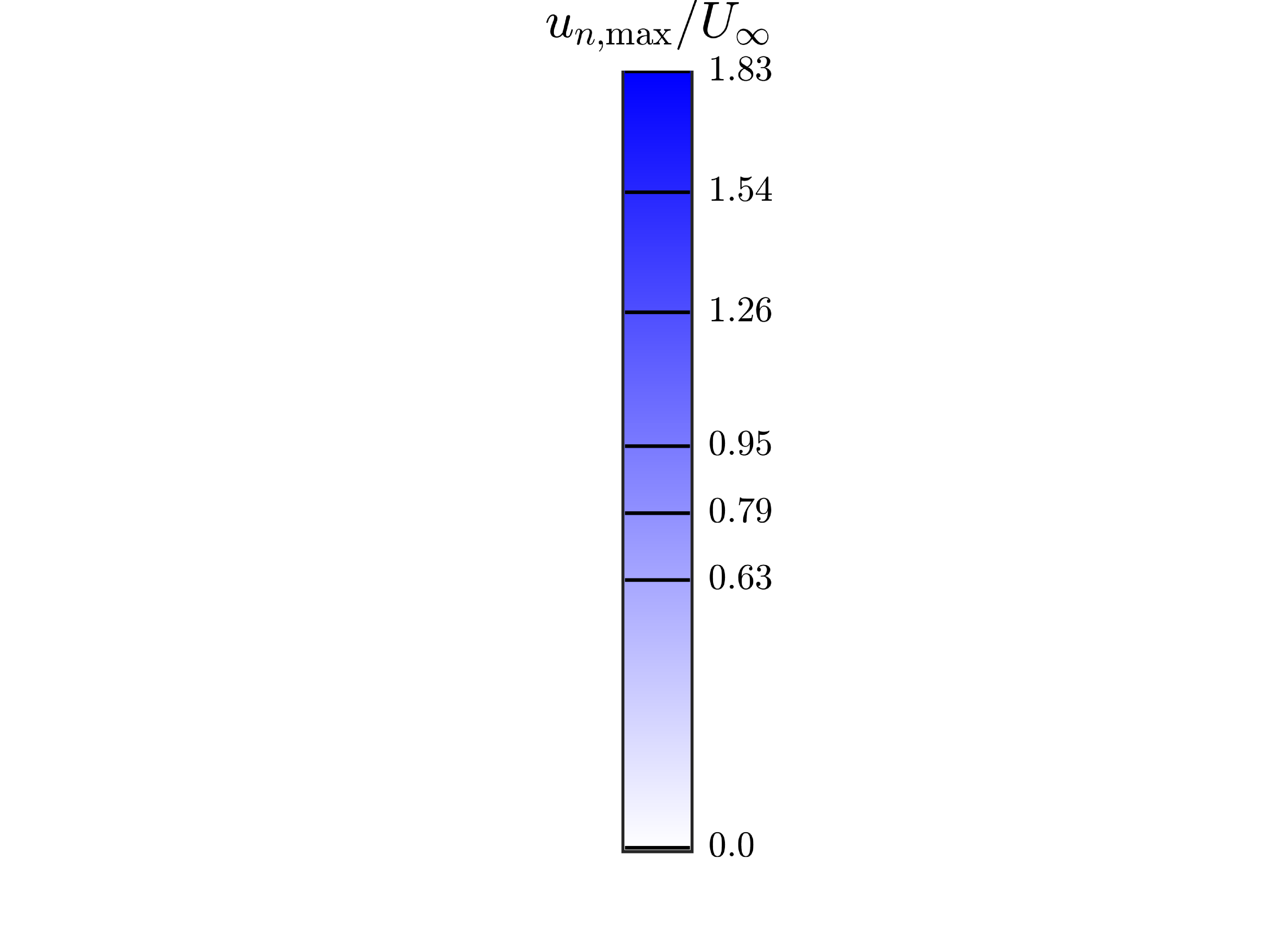}} 
		\caption{The correction term $S^2 C_\mu$ versus the lift and drag forces. The baseline values are indicated by \dashed ~and the controlled cases include pure blowing ($\bigcirc$), co-rotating ($\triangledown$), and counter-rotating ($\triangle$).}
		\label{fig:CoeffTheta}
\end{center}
\end{figure}

The aerodynamic forces modified by flow control over $2S C_\mu$ and $S^2 C_\mu$ are plotted in \figs \ref{fig:CoeffSwirl} and \ref{fig:CoeffTheta}, respectively.  These figures respectively show the symbols with shading based on the level of $u_{\theta,\max}/U_\infty$ and $u_{n,\max}/U_\infty$.  The inspection of these figures do not reveal a pronounced trend in the force data for the range of wall-normal momentum and swirl injections considered.  There is a wide scatter of the force data on the plots, without considering the whole composition of $C_\mu^*$ as we will observe below.  

For cases of actuation that only consider swirl with minimal or no wall-normal momentum ($S \gg 1$), the modified coefficient of momentum will be dominated by the last correction term; $C_\mu^* \rightarrow S^2 C_\mu$.  Such case can be of importance in modifying boundary layer flows, as seen in the study of drag reduction for turbulent channel flow using rotating discs by Ricco and Hahn [\citen{RiccoJFM13}].  Here, we do not place emphasis on the case of $S \gg 1$ because some level of wall-normal momentum is needed to penetrate the actuator input out of the boundary layer to achieve effective suppression of airfoil separation.  

\begin{figure}
\begin{center}
\includegraphics[trim={0 5cm 7cm 0},clip,width=0.62\textwidth]{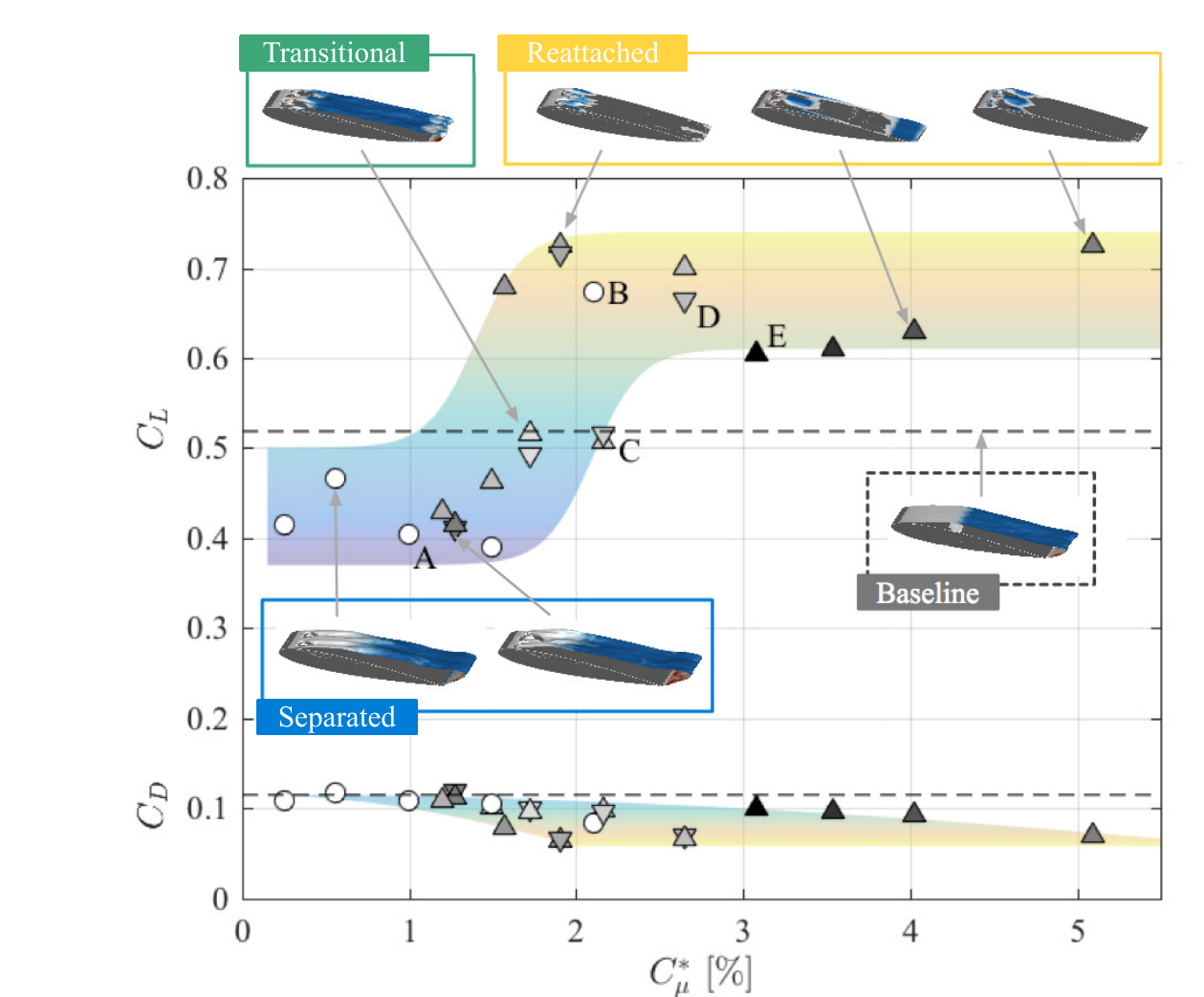} 
\raisebox{0.45 in}{\includegraphics[trim={8.5cm 0 6cm 0},clip,width=0.13\textwidth]{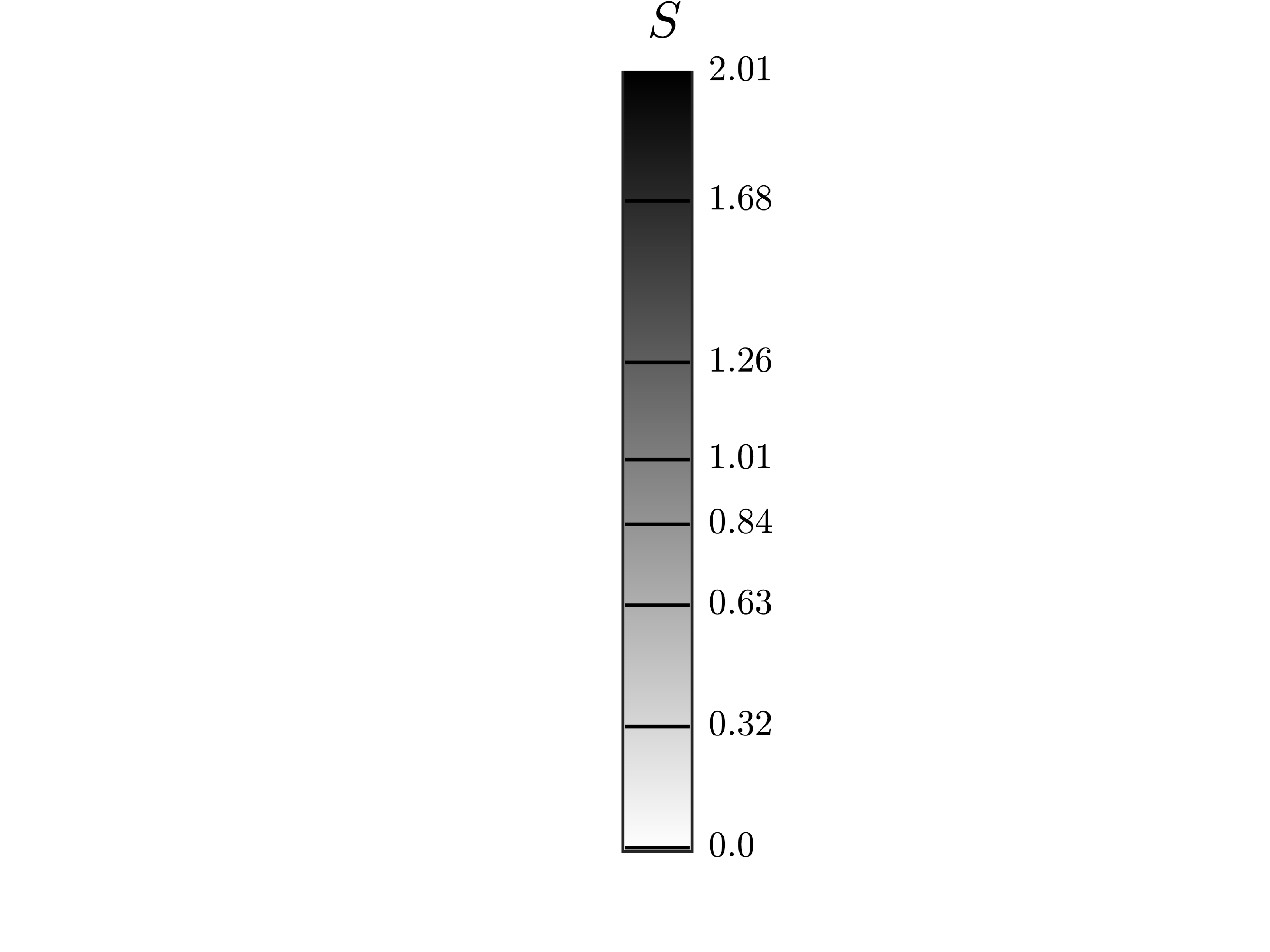}} 
\caption{The drag (bottom) and lift (top) forces versus the modified coefficient of momentum $C_\mu^* = (1+S)^2 C_\mu$.  The baseline values are indicated by \dashed ~and the controlled cases include pure blowing ($\bigcirc$), co-rotating ($\triangledown$), and counter-rotating ($\triangle$).  Inserts in the figure are representative time-averaged flows using $\overline{u}_x = 0$ iso-surface with Reynolds stress $\tau_{xy}$ overlaid in color (follows the visualization set-ups of \figs \ref{fig:baseline} and \ref{fig:A9_sepReyZ}).}
\label{fig:ForceA9_CT}
\end{center}
\end{figure}

Let us next examine whether the modified coefficient of momentum $C_\mu^*$ from \eq (\ref{eq:Cmustar2}) enables the quantification of the flow control effectiveness (lift enhancement and drag reduction) in terms from the combined forcing inputs of wall-normal momentum and azimuthal swirl.  Shown in \fig \ref{fig:ForceA9_CT} are the lift and drag forces plotted over $C_\mu^*$.  Unlike $C_\mu$ and the two correction terms, \eqs (\ref{eq:correct1}) and (\ref{eq:correct2}), visible trends emerge in the lift and drag response to flow control over $C_\mu^*$.  In \fig \ref{fig:ForceA9_CT}, we identify three broad classifications of the flow as highlighted by the colored shades in the background; namely they are (i) separated flow ($C_\mu^* \lesssim 1.5\%$; blue), (ii) transitional flow ($1.5\% \lesssim C_\mu^* \lesssim 2\%$; green), and (iii) reattached flow ($2\% \lesssim C_\mu^*$; yellow). Inserts in the figure are representative time-averaged flow visualizations using $\overline{u}_x = 0$ iso-surface with Reynolds stress $\tau_{xy}$ shown in color, following the visualization setups of \figs \ref{fig:baseline} and \ref{fig:A9_sepReyZ}.


For low values of $C_\mu^* \lesssim 1.5\%$, the flow remains separated as shown by the blue region in \fig \ref{fig:ForceA9_CT}.  The shown representative flows are similar to the flow shown above from case A (see \fig \ref{fig:A9_sepReyZ}), in which the time-averaged recirculation region is approximately the same size as the baseline case. The control input alters the shear layer emanating from the leading edge but is not sufficient of a change leading to reattachment.  For this low control input ($C_\mu^* \lesssim 1.5\%$), flow control in fact negatively impacts the flow resulting in lift decrease compared to the baseline case. 

The second region observed in \fig \ref{fig:ForceA9_CT} is the transitional flow cases for $1.5\% \lesssim C_\mu^* \lesssim 2\%$, where larger lift and lower drag values are starting to be achieved, as highlighted by the green shading in \fig \ref{fig:ForceA9_CT}. Around the representative case shown by case C, the combined control input modifies the separated flow but it remains separated behind the actuators. As seen from the variation in the lift and drag values in this region, the flow becomes sensitive to changes in the control input in this transitional region. The use of control input with the modified momentum coefficient of $1.5\% \lesssim C_\mu^* \lesssim 2\%$ exhibits varied effects on the flow and variation in the achieved lift values.  It can also be observed the onset of drag reduction starts over this region of $C_\mu^*$.

The third, right-most region, $C_\mu^* \gtrsim 2\%$, is reattached flow, highlighted in yellow in \fig \ref{fig:ForceA9_CT} and already discussed as case D in \fig \ref{fig:A9_sepReyZ}. In this region, the controlled flow is reattached downstream of the actuators. The reattached flow from control results in improved aerodynamic performance and some of the largest lift values attained in this study (increase of $\approx40\%$). All of the cases we examined with a modified coefficient of momentum greater than $2\%$ enables lift enhancement. Moreover, the corresponding drag force is significantly reduced due to the reattachment for most cases.  Since the maximum attainable increase in lift and reduction in drag is realized with fully reattached flow, simply further increasing $C_\mu^*$ does not continue increasing lift or reducing drag.  

Based on these observations of lift as a function of $C_\mu^*$, it can be said that a critical $C_\mu^*$ needed for reattachment of the flow is approximately $2\%$ such that maximal lift enhancement and drag reduction can be realized.  We do not expect additional benefit in aerodynamic enhancement by injecting actuator input beyond that critical value.  We also point out that while $C_\mu^*$ captures the overall trend in the enhancement of lift, overlaying the swirl value $S$ (shown by the shading of the symbols in \fig \ref{fig:ForceA9_CT}) can provide additional insights especially for control cases with higher $C_\mu^*$.

Through close examination of flow control results presented in \fig \ref{fig:ForceA9_CT}, it can be noticed that there are three cases with $C_\mu^* > 3\%$ that achieve lift enhancement but with somewhat lower increase in lift and reduction in drag.  These cases are where $S > 1.5$; specifically $C_\mu^* = 3.08$ (case E, $S = 2.51$), $C_\mu^* = 3.54$ ($S = 2.01$), and $C_\mu^* = 4.03$ ($S = 1.68$), shown by the dark shaded symbols.  For these three cases with large swirl inputs, variations of time-averaged flow suggest that the actuation jets are able to reattach the flow at first glance.  As these three cases exhibit similar characteristics of the flow, let us examine case E in further detail.

\begin{figure}
	\begin{center}
	\includegraphics[width=0.8\textwidth]{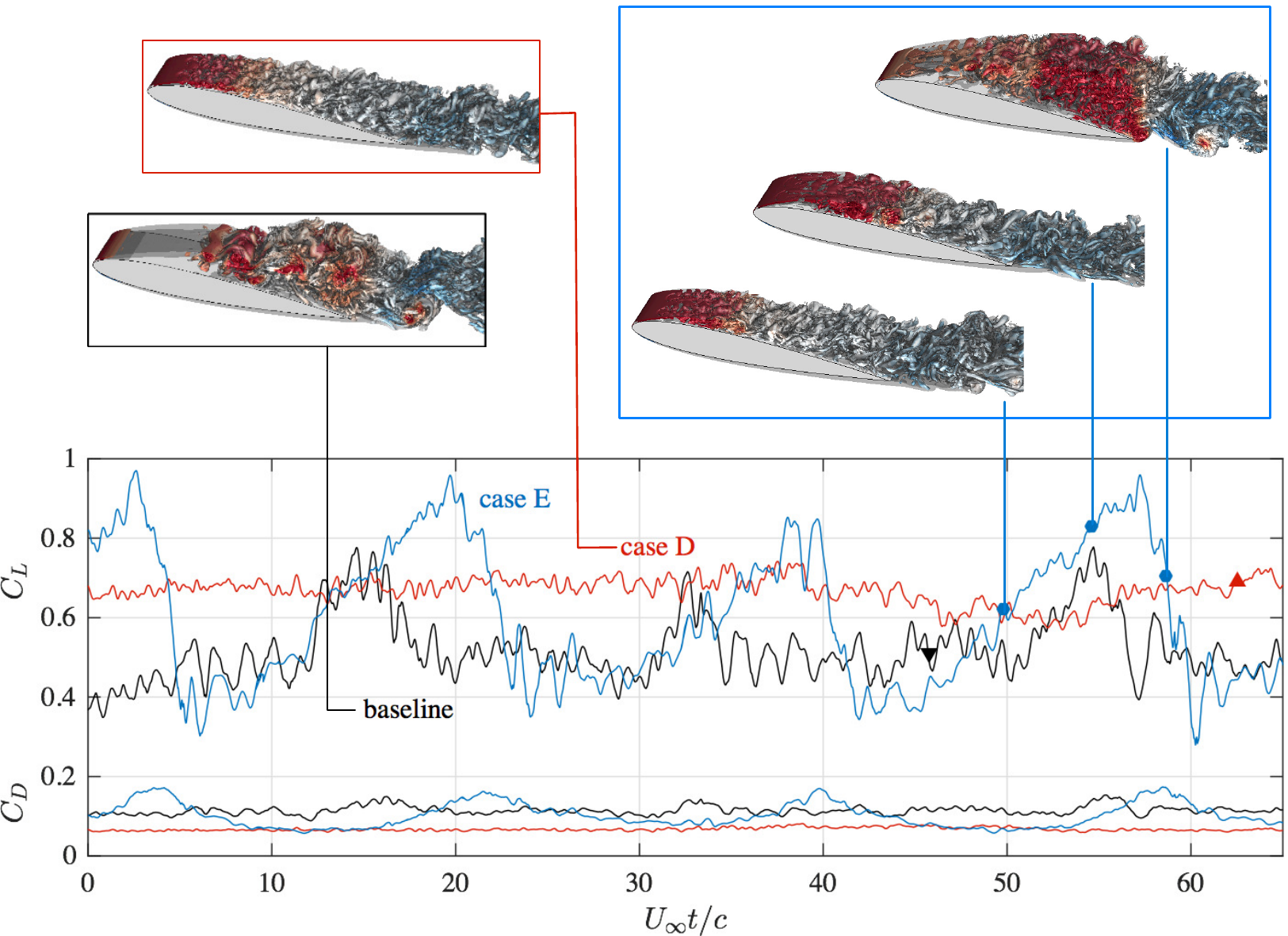} 
	\caption{Time history of lift and drag for baseline ($-$), case D ({\color{red}$-$}), and case E ({\color{blue}$-$}). Representative instantaneous flow fields are visualized with the $Q$-criterion isosurface colored with pressure, following the same setup in \fig \ref{fig:baseline}.  Symbols on the lift curves indicate the time of visualization.}
	\label{fig:ForceTime}
	\end{center}
\end{figure}

We compare the flow fields and lift histories from case E and those from the baseline case and case D (that successfully reattaches the flow) in \fig \ref{fig:ForceTime} to explain why the three large-swirl cases experience lower lift increase.  Case D achieves larger lift increase and significant drag reduction.  Moreover, the fluctuations in lift and drag are significantly reduced from the baseline oscillations. This suppression of large-scale unsteadiness is realized by breaking up large-scale spanwise vortical structures (associated with strong pressure cores), which are present in the baseline case (see baseline case in \fig \ref{fig:ForceTime}). 

\begin{table} 
\centering
\begin {tabular}{l c c} \hline
Case 	& $\text{RMS}(C_D)$ 	&$\text{RMS} (C_L)$ \\  \hline\hline	
Baseline	& 0.011				& 0.083			\\ \hline
A  		& 0.006				& 0.041				\\ 
B		& 0.001				& 0.011			    \\
C  		& 0.007				& 0.043			\\ 
D  		& 0.006				& 0.044			\\
E  		& 0.029				& 0.167		    \\ \hline
\end{tabular}
\caption{Root mean square of the aerodynamic forces for the representative cases.}
\label{tbl:A9_RMS}
\end{table}

The use of high-swirl input ($S = 2.51$) in case E accentuates the low-frequency lift fluctuation present in the baseline flow.  Case E produces large force fluctuations in aerodynamic forces, which can be seen from the RMS values in Table \ref{tbl:A9_RMS}.  For case E, we find that the large input of angular momentum (swirl) in relationship to wall-normal momentum results in the generation of the low-frequency unsteady flow feature reminiscent of dynamic stall.  The flow appears as mostly attached over the airfoil as shown by the instantaneous flow fields at $t \approx 50$ and $54$ in \fig \ref{fig:ForceTime}.  Over time, there is a slow build up vortical structures from the leading edge, yielding a low-pressure distribution to grow as illustrated by the red region over the wing.  Once this structure extends over the majority of the wing, we see a massive detachment of the accumulated structure which causes the airfoil to loose the enhancement in lift. This dynamic process is similar to the unsteady flow behavior known as dynamic stall for pitching wings [\citen{Carr:JA88, Garmann:PF11}], although the wing in our study is not undergoing a pitching maneuver. Dynamic stall process experiences increased influx of surface vorticity from the acceleration of the wing as Eq. \ref{eq:vortFlux} reveals. In the current study, the added vorticity input to the flow from swirl contributes greatly to the near-wall vorticity dynamics to significantly modify the unsteady features in these high-swirl cases.  Since the detachment takes place over a shorter period of time, the time-averaged flow appears as attached as seen in \fig \ref{fig:ForceA9_CT}.

These low-frequency fluctuations are known to be triggered by other actuation mechanisms also.  Through wind-tunnel experiments, Bernardini et al.~[\citen{Bernardini:AIAAJ16}] report that similar large unsteadiness for flow over post-stall airfoils can be caused by moderate-amplitude actuation inputs, including acoustic excitation as well as steady and pulsed actuation jets.  They also observe separated flow responding to actuation input in a manner similar to dynamic stall.  Their conditionally-averaged flow visualizations reveal the bursting and reattachment of the separation bubble resulting in large-scale force fluctuations.  

Based on the present observations, we can deduce that the modified coefficient of momentum $C_\mu^*$ can capture the overall lift enhancement and drag reduction trend by incorporating the swirl coefficient $S$, as captured in \fig \ref{fig:ForceA9_CT}.  For the present setup, we have observed that flow control settings with $C_\mu^* \gtrsim 2\%$ enable lift enhancement.  We however also note that application of large swirl input should be treated with care since it can cause the flow to exhibit large-scale unsteadiness.  While the use of $C_\mu^*$ over a range of $S$ values in assessing the effectiveness in lift enhancement appears to collapse the lift and drag response, we should note that we have incorporated the influence of $S$ only in a linear manner into the characteristic jet velocity, which might have implicitly assumed small $S$ values.  The utilization of much larger $S$ value than what is considered in the current study may not be as effective nor might it show a collapse of the flow control results; which may share similarity with how we can detrimentally overpower separated flow with high $C_\mu$ inputs.  From this investigation and knowledge gained from past studies cited throughout this paper, we note that the use of wall-normal momentum injection appears to be the primary control input with angular momentum (swirl) injection serving as a companion control input that can enhance the overall effectiveness of flow control.  This notion is also consistent with the findings gained from the use of the modified coefficient of momentum, \eq (\ref{eq:Cmustar1}), to assess the effectiveness of wall-normal and angular momentum injections to achieve lift enhancement and drag reduction.


\section{Conclusions}

This study examined the influence of steady wall-normal and angular (swirl) momentum injections on suppressing flow separation over a NACA 0012 airfoil at $Re = 23,000$ and $\alpha = 9^\circ$ by performing a number of LES computations. The baseline flow at this condition exhibits massive separation as the shear layer detaches from the leading edge and rolls up into spanwise vortices, which later develop into a turbulent wake downstream. Flow control inputs are introduced to increase lift and reduce drag by eliminating flow separation that covers the entire suction surface of the airfoil.  Instead of replicating a specific type of actuators, we introduced fundamental control inputs by prescribing wall-normal and swirling velocity profiles over circular ports near the natural separation location.  These two fundamental actuation inputs were selected in this study due to the prevalence of utilizing momentum and vorticity (as well as mass) injections in many flow control studies.  The responses of the flow field and the aerodynamic forces to varied level of control inputs were studied systematically through a number of flow control cases. 

The instantaneous and time-averaged flow fields under the influence of only the wall-normal injection revealed that such actuation can reattach the flow with coefficient of momentum of $C_\mu \gtrsim 2\%$.  With sufficient wall-normal momentum input, the lift is increased significantly by eliminating the loss induced by the large separated region.  The addition of angular momentum (swirling jet) to the actuation enabled effective separation suppression with notably lower level of coefficient of momentum.  The added vorticity (swirling) inputs further enhanced mixing between the low-momentum fluid near the airfoil and the high-momentum freestream, overcoming the adverse pressure gradient.  To examine the effect of flow control on the separated flow, the vortical structure, surface vorticity flux, and the Reynolds stress distributions were closely examined.  

The effects of wall-normal momentum and angular momentum on lift enhancement and drag reduction were captured through a single consolidated control input parameter.  This was achieved by identifying a characteristic velocity $u_j^*  = (1+S) u_{n,\max}$ for the swirling actuator jet by incorporating the non-dimensional jet swirl number $S$; a ratio of the angular momentum to the wall-normal momentum.  For the present control setting, the swirl ratio reduces to the azimuthal to wall-normal velocity ratio, $S = \kappa \frac{u_{\theta,\max}}{u_{n,\max}}$. Based on the characteristic swirling jet velocity $u_j^* $, we derived a modified coefficient of momentum $C_\mu^*$, which is related to the traditional coefficient $C_\mu$ by $C_\mu^* = (1+S)^2 C_\mu$.  

Using this modified coefficient of momentum we have been able to consolidate the behavior of modified aerodynamic forces from flow control and classify different flow control cases as: (1) separated flow for $ C_\mu^* \lesssim 1.5\%$, (2) transitional flow for $1.5\% \lesssim C_\mu^* \lesssim 2\%$, and (3) reattached flow for $2 \% \lesssim C_\mu^*$.  For small values of modified momentum coefficient $C_\mu^* \lesssim 1.5\%$, lift decreases and the flow remains separated. For these cases with small inputs of control, the size of the separation region remains similar to that of the baseline case.  A drastic increase in lift forces is observed when the reverse flow region begins to diminish as $C_\mu^*$ exceeds $1.5\%$.  Once the control input surpasses $C_\mu^* \gtrsim 2\%$, flow is reattached, leading to improvements in the aerodynamic forces.  Based on these observations, the currently considered flow should require $C_\mu^* \gtrsim 2\%$ to ensure complete reattachment and aerodynamic force enhancement.  The influence of large-swirl values was also studied.  A cautionary note was provided not to use too large of a swirl ratio ($S > 1.5$) in controlling the flow, in which case the flow exhibits lift enhancement but with unsteady behavior reminiscent of dynamic stall.

We have discussed how the combined use of steady wall-normal and angular momentum injections (swirling jets) can effectively reattach separated flow over a canonical NACA0012 airfoil.  Although this study focused on a chosen angle of attack and Reynolds number, the present control strategy to influence separated flow should be effective for other separated flows at similar Reynolds numbers.  There are open questions on assessing the influence of freestream turbulence and higher Reynolds number flows on the effective characterization of flow control.  Nonetheless, we have observed that the control technique and quantification approach discussed above are also effective for wider spanwise spacing of actuators and different angles of attack, which are not presented  to maintain some brevity of this paper [\citen{MundayPhd17}].  The present approach to incorporate the influence of swirl injection to determine the characteristic actuation velocity and, in turn, evaluate the modified coefficient of momentum is general in nature, which we suspect is applicable to other types of flows.

\section*{Acknowledgments}

This research was supported by the U.S. Air Force Office of Scientific Research (Award Number FA9550-13-1-0183) and the U.S. Office of Naval Research (Award Number N00014-16-1-2443).  The computations were performed on the large-scale computing clusters made available through the High Performance Computing Modernization Program at the U.S. Department of Defense. PMM also thanks the Lockheed--Martin graduate fellowship.

\appendix

\section{List of flow control cases}
\label{app:A9}
Table \ref{tbl:A9_Cases} summarizes all LES cases of separation control performed for this study. The control setting with the associated non-dimensional input parameters and resulting drag and lift forces are tabulated.

\begin{table}  
\centering
{\small \begin{tabular}{l c c c c c c c c c c} \hline
Case 	& $\frac{u_{n,\max}}{U_\infty}$ 	 &$\frac{u_{\theta,\max}}{U_\infty}$ & $C_\mu~[\%]$ & $S$ & $C_\mu^*~[\%]$  & Rot.~dir.  & $C_D$&$ C_L$ & $\text{RMS}(C_D)$ & $\text{RMS}(C_L)$ \\  \hline\hline	
Baseline& --		& --			& --        & --        & --    & -- 	 & 0.115	& 0.519 & 0.011 & 0.083 \\ \hline
--  & 0             & 5.04 			& 0 &$\infty$ & $\infty$ & CTR 	& 0.118	& 0.552 & 0.014 & 0.100 \\
		
-- & 0.631			& 0				& 0.25 & 0 & 0.25		& --		& 0.108	&	0.416	& 0.006 & 0.046 \\ 
-- & 0.631			& 1.26			&0.25 & 1.26 & 1.27		& ROT 	 	& 0.120	&	0.411	& 0.009 & 0.060\\ 
-- & 0.631			& 1.26			&0.25 & 1.26 & 1.27		& CTR	 	& 0.112	&	0.414	& 0.008 & 0.061 \\ 
E  & 0.631          & 2.52          &0.25 & 2.52 & 3.08     & CTR       & 0.010 &   0.605  & 0.029 & 0.167 \\

-- & 0.789			& 0.95			& 0.391 & 0.755  & 1.20		& CTR	 	& 0.108	&	0.429 & 0.007 & 0.038\\ 
-- & 0.789			& 1.26			& 0.391 & 1.006  & 1.57		& CTR 	 	& 0.078	&	0.679 & $--$ & $--$\\ 
-- & 0.789			& 2.52			& 0.391 & 2.013  & 3.54		& CTR	 	& 0.097	&	0.609 & $--$ & $--$\\ 

-- & 0.946			& 0				& 0.563 & 0 & 0.56		  & --		& 0.118	&	0.467	& 0.011 & 0.088\\ 
-- & 0.946			& 0.946			& 0.563 & 0.629 & 1.49		& CTR 	 	& 0.101	&	0.463 & 0.009 & 0.052\\ 
-- & 0.946			& 1.26			& 0.563 & 0.839 & 1.90		& ROT	 	& 0.065	&	0.716 & 0.002 & 0.018\\ 
-- & 0.946			& 1.26			& 0.563 & 0.839 & 1.90		& CTR	 	& 0.064	&	0.726 & 0.002 & 0.014\\ 
-- & 0.946			& 2.52			& 0.563 & 1.677 & 4.03		& CTR 	 	& 0.092	&	0.628 & 0.031 & 0.178\\ 

A  & 1.26			& 0				& 1.00 & 0 & 1.00		& --		 	& 0.108	&	0.403 & 0.006 & 0.041\\ 
-- & 1.26			& 0.631			& 1.00 & 0.314 & 1.73		& ROT	 	& 0.099	&	0.493 & 0.006 & 0.047\\ 
-- & 1.26			& 0.631			& 1.00 & 0.314 & 1.73		& CTR 	 	& 0.096	&	0.510 & 0.010 & 0.067\\ 
C  & 1.26			& 0.95			& 1.00 & 0.472 & 2.17		& ROT 	 	& 0.096	&	0.516 & 0.007 & 0.043\\ 
-- & 1.26			& 0.95			& 1.00 & 0.472 & 2.17		& CTR 	 	& 0.097	&	0.507 & 0.012 & 0.081\\ 
D  & 1.26			& 1.26			& 1.00 & 0.63 & 2.65		& ROT	 	& 0.069	&	0.665	& 0.006 & 0.044 \\ 
-- & 1.26			& 1.26			& 1.00 & 0.63 & 2.65		& CTR	 	& 0.067	&	0.699 & 0.002 & 0.028\\ 
-- & 1.26			& 2.523			& 1.00 & 1.26& 5.09		& CTR	 	& 0.069	&	0.725 & 0.002 & 0.025\\ 

-- & 1.545			& 0			& 1.50 & 0 & 1.50		& -- 	 		& 0.106	&	0.391 & 0.009 & 0.061 \\ 
B  & 1.829			& 0			& 2.10 & 0& 2.10		& --	 		& 0.084	&	0.673 & 0.001 & 0.011 \\ \hline		
	\end{tabular}}
\caption {Compilation of the non-dimensional control input parameters and the resulting drag and lift coefficients including their RMS values. For cases with $--$, data is unavailbe.}
\label{tbl:A9_Cases}
\end{table}

\bibliographystyle{aiaa}
\bibliography{refs.bib}

\end{document}